 \documentclass[final,5p,times,twocolumn]{elsarticle}
 \usepackage{moresize}
 \usepackage{appendix}
 \usepackage{caption}
 \usepackage{adjustbox}
 \usepackage[utf8]{inputenc}
 \usepackage{graphicx}
 \usepackage{supertabular}
 \usepackage{xcolor}
 \usepackage{paracol}
 \usepackage{lscape}
 \usepackage{amsmath}
 \usepackage{amsfonts}
 \usepackage{amssymb}
 \usepackage{natbib}

\usepackage{amssymb}
\usepackage{lipsum}
\usepackage{amsmath}
\usepackage{tabularx,ragged2e}
\usepackage{booktabs}
\usepackage{longtable}
\usepackage{multicol}
\usepackage{array, booktabs, makecell}
\usepackage{siunitx, mhchem}

\journal{}

\begin{document}

\begin{frontmatter}



\title{Effect of Alloying on Intrinsic Ductility in WTaCrV High Entropy Alloys}


\author[first]{Akshay Korpe}
\affiliation[first]{organization={School of Mechanical and Automotive Engineering, Clemson University}, 
            city={Clemson},
            postcode={29634}, 
            state={South Carolina},
            country={United States of America}}
\author[third]{Osman El-Atwani}
\affiliation[third]{organization={Pacific Northwest National Laboratory},
            addressline={}, 
            city={Richland},
            postcode={29634}, 
            state={Washington State},
            country={United States of America}}
\author[first,second]{Enrique Martinez Saez}
\affiliation[second]{organization={Department of Materials Science and Engineering, Clemson University},
            city={Clemson},
            postcode={29634}, 
            state={South Carolina},
            country={United States of America}}
            
\begin{abstract}
Tungsten (W) exhibits desirable properties for extreme applications, such as the divertor in magnetic fusion reactors, but its practicality remains limited due to poor formability and insufficient irradiation resistance. In this work, we study the intrinsic ductility of body-centered cubic WTaCrV based high entropy alloys (HEAs), which are known to exhibit excellent irradiation resistance. The ductility evaluations are carried out using a criterion based on the competition between the critical stress intensity factors for emission ($K_{Ie}$) and cleavage ($K_{Ic}$) in the \{110\} slip planes and \{110\} crack planes, which are evaluated within the linear elastic fracture mechanics framework and computed using density functional theory calculations. The results suggest that increasing the alloying concentrations of V and reducing the concentrations of W can significantly improve the ductility in these HEAs. The elastic anisotropy for these HEAs is analyzed using the Zener anisotropy ratio and its correlation with the concentration of W in the alloys is studied. Results indicate that these alloys tend to be fairly isotropic independently from the concentration of W in them. The computed data for the elastic constants of these HEAs is also compared against the available experimental data. The results are in good agreement, hence validating the robustness and accuracy of the computational methods. Multiple phenomenological ductility metrics were also computed and analyzed against the analytical model. The results suggest that these models may have higher computational efficiency due to less number of parameters required for their computation. Some models, like the surrogate D parameter and the Pugh ratio, show a good correlation with the Rice model. The potential of these empirical models to serve as surrogate screening models for optimizing the compositional space is also discussed. 
\end{abstract}




\end{frontmatter}




\section{Introduction}
\label{introduction}

In the realm of magnetic fusion reactors, plasma-facing materials (PFMs), such as the divertor in tokamak devices, are subjected to extreme environments during operation with long exposures (\(>\) $\mathrm{10^7s})$ to high temperatures (\(>\) 1000 K) and ionized and energetic deuterium (D) and tritium (T), helium (He) ash fluxes (\(>\) $\mathrm{10^{24} m^{-2} s^{-1}})$ and neutrons \cite{Baldwin_2008}. The materials used for these applications therefore need to possess exemplary thermo-mechanical properties such as high melting points and good strength retention at elevated temperatures as well as excellent irradiation resistance including low sputter erosion and low tritium retention \cite{doi:10.1126/sciadv.aav2002}. In recent years, a new class of tungsten (W) based high entropy alloys (HEAs) of the form WTaCrV has been studied to show outstanding irradiation resistance that makes them potential candidates for PFM applications. There is good computational data backed by experimental verification that shows the viability of these alloys in environments of extreme irradiation, which tests these materials at elevated temperatures (1073 K) and irradiation up to 8 dpa showing good retention of mechanical properties after exposure \cite{doi:10.1126/sciadv.aav2002}. However, the ductility of these HEAs is not well understood, with not many studies available. In the few experimental works in the literature, these WTaCrV based HEAs show poor ductility (1-2 \% fracture strains) at room temperatures \cite{WASEEM201887}. This problem of poor ductility as well as the relatively sparse understanding of the underlying ductility mechanisms has put forward new challenges both in computational and experimental research pertaining to the viability of these materials in areas of nuclear fusion reactors. The ductility of a material is a widely accepted surrogate indicator of its potential to achieve high fracture toughness. It is also an important property to predict the formability and manufacturing potential of metallic alloys and hence there is an urgent need to study the ductility of these HEAs. \par
Body-centered cubic (BCC) refractory HEAs are generally known to be intrinsically brittle at room temperature \cite{Tsuru2024-bc,LI2021100777}. Moreover, many BCC refractory metals (V, Ta, Cr, Mo, W) have been observed to show a sharp brittle-to-ductile transition with increasing temperature and are known to be brittle at lower temperatures \cite{GUMBSCH2003304,Savitskii1970,semchishen1962molybdenum}. This implies that the brittleness of these BCC HEAs is not a property solely to be attributed to HEAs and is a characteristic feature of many other BCC refractory metals. These alloys do not show a clear correlation between the yield strengths and ductility as seen in experiments performed to study ductility under compressive loading. BCC alloys like equimolar VNbMoTaW (1246 MPa at 296 K with 0.5 \%\ peak strain) and NbMoTaW (1058 MPa at 296 K with 1.5 \%\ peak strain) show very low ductility at relatively high yield strengths at ambient temperatures  \cite{SENKOV2011698} whereas similar alloys like MoNbTaTi (1800 MPa at 293 K with ~25 \%\ peak strain) and WNbTaTi (1800 MPa at 293 K with ~20 \%\ peak strain) show high ductility at room temperature with high yield strengths \cite{COURY201966}. Also, the brittle-to-ductile transition temperatures (BDTT) are quite sharp and hence cannot be explained completely by a dislocation plasticity model, which would predict the transition to be smooth \cite{GUMBSCH2003304,Hartmaier1999145}. Attributable to these anomalous behavior exhibited by BCC refractory HEAs, new models for ductility prediction have been the topic of contemporary research. \par
Most of the available experimental datasets reporting ductility of BCC HEAs are for compressive loading \cite{COURY201966,SENKOV2011698,WASEEM201887}, at least partly for testing the manufacturing potential in the form of formability. Experimental evidence indicates that the mode of failure in compression is mode I (tensile failure) \cite{SENKOV2011698}. Many ideas have been put forward to understand the underlying mechanisms that distinguish between ductile and brittle responses in materials. One of the most widely used frameworks explains intrinsic ductile behavior in terms of atomic structure at an atomistically sharp crack, corresponding to the competition between dislocation emission and cleaving \cite{R_Thompson,Ductile_and_brittle_crack-tip_response}. This theory was further developed by Rice and Thomson into continuum models for modeling the behavior of pre-cracked bodies under applied loads \cite{doi:10.1080/14786437408213555}. This theory is in accordance with experimental studies of crack tip behavior using transmission electron microscopy (TEM), which report extensive dislocation emission and blunting in Nb, which is known to be intrinsically ductile and very few dislocation emissions in W, which is known to be a brittle metal at room temperatures \cite{OHR19851}. These observations are also consistent with the BDTT behavior of these BCC HEAs. It has to be noted however, that there are experimental works on BCC HEAs that attribute intrinsic brittleness to inter-granular fracture due to formation of brittle oxides and nitrides at the grain boundaries without significant crack tip plasticity \cite{ZOU201795}. \par
In this work, we rely on Rice's model using Peierls framework into dislocation nucleation \cite{RICE1992239} along with the concept of crack propagation due to Griffith's cleavage \cite{Griffith1921163} to analyze the ductility of the previously mentioned WTaCrV based HEAs. We make use of the postulated hypothesis by Mak et al. \cite{MAK2021104389} stating that a material is intrinsically ductile if dislocation emission occurring at a sharp crack tip mode I stress intensity factor ($K_{Ie})$ occurs prior to the cleavage fracture at crack tip mode I stress intensity factor $(K_{Ic})$. The assumption to consider only mode I loading is based on experimental work by Senkov et al. \cite{SENKOV2011698}. The idea is that dislocation emission blunts the sharp crack tip and is the enabling mechanism for the subsequent onset of ductile failure mechanisms (void nucleation, growth, and coalescence ahead of the crack). Otherwise, a sharp crack remains sharp and propagates easily. \par
This framework allows for dislocation plasticity around the crack tip, leaving the tip to be still sharp providing extra energy dissipation. This allows for the macroscopic toughness of a material to be much higher than the Griffith \cite{Ductile_and_brittle_crack-tip_response} cleavage value (but still lower than the value achievable by ductile failure mechanisms) and for a sharp transition in the toughness due to the crack tip behavior. This work also neglects the effects of twinning and only considers full dislocation emissions which is in agreement with experimental observations \cite{OHR19851}.    \par
Based on this postulate, we have made a systematic theoretical analysis of the intrinsic ductility of WTaCrV based BCC refractory HEAs for varying concentrations of all four constituent elements and assess their correlation to the calculated ductility. The displacement and stress fields were calculated using linear elastic fracture mechanics (LEFM) for a crack tip in an isotropic elastic medium \cite{Ting}. The displacement and stress fields are summarized in \ref{appendix A1}. Out of these alloys, $\mathrm{W_{37}Ta_{37}Cr_{15}V_{11}}$ has been previously synthesized and studied for properties such as the enthalpy of mixing $\mathrm{(\Delta H_{mix})}$ and the elastic constants, which are used as a reference to validate the computational results \cite{doi:10.1126/sciadv.aav2002,QIN2024108384}. The required parameters are determined computationally using first-principles density functional theory (DFT) calculations. \par
The remainder of this paper is structured as follows. Section \ref{Rice theory} presents a brief summary of Rice theory that is used for quantifying the intrinsic ductility of the BCC refractory HEAs of interest, as well as a detailed explanation of the DFT methodology employed to compute all the relevant parameters. In Section \ref{sec:phenomeno}, we have a detailed description of the main results of this research with additional discussion. We compare our findings with experimental data and discuss factors that may be responsible for any disagreement. We also discuss the correlation of the concentration of each element with the material properties of these alloys as well as study their effects on the intrinsic ductility of these HEAs. We end up the study highlighting the main conclusions in Section \ref{sec:conclusions}.

\section{Methodology}
\subsection{Rice theory for crack propagation and dislocation nucleation at a sharp crack tip}
\label{Rice theory}

The crack tip is considered to be atomistically sharp in an infinite anisotropic, elastic medium under plane strain (2-dimensional medium) and mode I (tensile) loading. We have two competing mechanisms under consideration, crack propagation due to Griffith's cleavage \cite{Griffith1921163} and dislocation emission at the crack tip \cite{RICE1992239}. LEFM for 2-d anisotropic elastic medium \cite{Ting} has been used to calculate the displacement and stress fields, which is summarized in Appendix A1. This framework is used in tandem with first-principles DFT approaches to compute the parameters required to calculate the stress intensity factors for crack propagation ($\mathrm{K_{Ic}})$ and dislocation emission ($\mathrm{K_{Ie}})$. \par
\subsubsection{Geometry}
\label{Geometry}
\begin{figure}
    \centering
    \includegraphics[width=3in]{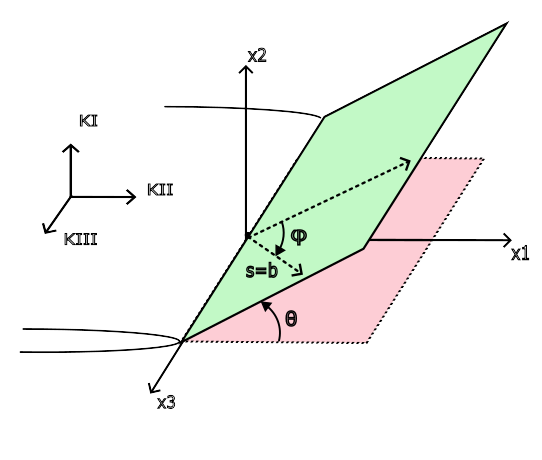}
    \caption{A schematic diagram representing the geometry of the crack systems. The red plane represents the crack plane and the green plane represents the slip plane. The angles $\mathrm{\theta}$ is the angle between the crack plane and the slip plane .Angle $\mathrm{\phi}$ is the angle between the projection of the direction of crack tip propagation in the slip plane with the burgers vector \textbf{b} in the slip plane. Vector \textbf{s} is the unit vector defined in sec \ref{Geometry} and is equivalent to burgers vector \textbf{b}. }
    \label{Crack png}
\end{figure}

The different coordinate systems used are cartesian $\mathrm{(x_{1},x_{2}, x_{3})}$ and cylindrical coordinates $\mathrm{(r,\theta,x_{3})}$. $\mathrm{x_1}$ is the direction of the crack tip propagation, $\mathrm{x_2}$ is the direction normal to the crack plane and $\mathrm{x_3}$ is the direction parallel to the crack front. The slip plane makes an angle $\mathrm{\theta}$ with the crack plane and the slip direction \textbf{s} makes an angle $\mathrm{\phi}$ with the crack tip direction projection on the slip plane (see Fig.~\ref{Crack png}). The slip direction is the direction of the burgers vector and is characterized in the slip plane by the unit vector \textbf{s}($\mathrm{\phi}$) = $\mathrm{[cos{\phi}, 0, sin{\phi}]^T}$. It is to be noted to avoid any confusion that this unit vector is defined in an orthogonal basis that is not the same as the basis defined in Fig.  \ref{Crack png} but in an orthogonal system obtained rotating the $\mathrm{x_{1}-x_{2}-x_{3}}$ basis from Fig. \ref{Crack png} by an angle $\mathrm{\theta}$ around the $\mathrm{x_{3}}$ axis. The rotation matrix Q (defined in the following section) is used to relate the two orthogonal basis sets to each other.\par

\subsubsection{Stress intensity factors and ductility prediction metric D}

\label{Stress intensity factors and ductility metric D}
The energy release rate given by Irwin for a sharp crack tip to propagate in an anisotropic elastic medium for plane strain conditions is given by \cite{SUN19941905} : 
\begin{equation}
    \mathrm{G = \textbf{K}^{T} \Lambda \textbf{K}}
\label{Equation 1}
\end{equation}
$\mathrm{\textbf{K} = [K_{II},K_{I},K_{III}]^T}$ is the external loading, where $\mathrm{(K_\alpha ~for~ \alpha = II, I , III)}$ are the stress intensity factors for mode II (shear in-plane), mode I (tensile), and mode III (shear anti-plane) loading, respectively. $\boldsymbol{\Lambda}$ is a matrix defined in the Stroh formalism that contains information about the crack orientation in an anisotropic medium (see \ref{appendix A3}). Griffith condition for pure mode I loading is given by Rice using Peierls framework \cite{RICE1992239}:
\begin{equation}
    \mathrm{G = 2\gamma_s}
    \label{Equation 2}
\end{equation}
\noindent where $\mathrm{G}$ is the free surface energy of the crack and $\mathrm{\gamma_{s}}$ is the free surface energy. Using equations (\ref{Equation 1}) and (\ref{Equation 2}), 
 for pure mode I loading we get the following expression for the stress intensity factor under pure mode I loading :
 \begin{equation}
     \mathrm{K_{Ic} = \sqrt{\frac{2\gamma_{s}}{\boldsymbol{\Lambda}_{22}}}}
     \label{Equation 23}
 \end{equation}
 
 The Rice model \cite{RICE1992239} for linear-elastic, isotropic materials assumes that the displacement field around the crack tip is related to the shear stress and atomic displacement in a periodic manner. Slip displacement takes place on addition of load, which results in dislocation nucleation at the point of maximum lattice potential. The model treats the slip process as a pure shear process and uses the J-integral method to derive the following condition of instability for a complete lattice dislocation to nucleate :
 \begin{equation}
     \mathrm{G = \gamma_{usf}}
     \label{Equation 4}
 \end{equation}
Here, $\mathrm{\gamma_{usf}}$ is the unstable stacking fault (USF) energy which coincides with the maximum lattice potential. This is under the assumption that the slip is constrained to a straight slip path \textbf{s} without considering any dilational displacement perpendicular to the slip plane. The implication of this assumption is that the $\mathrm{\gamma_{usf}}$ used is the unrelaxed USF after the displacement in the in-plane directions. If the modeling takes into account the effects of tension-coupling, there will be a reduction in the critical load for dislocation nucleation. In order to account for this reduction, Sun and Beltz used an approximation that involves using the relaxed value of $\mathrm{\gamma_{usf}}$ instead of the unrelaxed values that give us equation (\ref{Equation 4}) \cite{SUN19941905}. The generalized expression for the nucleation condition of a co-planar crack $\mathrm{(\theta = 0)}$ in an anisotropic, linear-elastic medium is given by \cite{SUN19941905} :
\begin{equation}
    \mathrm{\frac{(\textbf{s}(\phi)\textbf{K}_{e})^2}{\textbf{s}^{T}(\phi)\boldsymbol{\Lambda}^{-1}\textbf{s}(\phi)} = \gamma_{usf}}
    \label{Equation 5}
\end{equation}
The numerator in equation (\ref{Equation 5}) can hence be thought of as a general case between stress intensity factors along the resolved shear stress in the slip direction \textbf{s} and $\mathrm{\textbf{K}_{e}}$ is a generalized stress intensity factor at emission at crack tip. Some special cases of loading and slip conditions are discussed here. For pure mode II loading, the emerging dislocation is an edge dislocation in the $\mathrm{x_{1}}$ direction $\mathrm{\phi = 0}$. Similarly, for pure mode III loading, the resulting dislocation is a screw dislocation in the $\mathrm{x_{3}}$ direction $\mathrm{\phi = \pi/2}$. The denominator of the equation has the term $\boldsymbol{\Lambda}^{-1}$ that contains information about crack orientation and slip direction. \par
We use Rice model \cite{RICE1992239} to determine the ``exact" criteria for dislocation nucleation in an anisotropic linear elastic material for non-coplanar crack and slip planes under a generalized set of stress intensity factors using numerical methods \cite{SUN19941905} :
\begin{equation}
   \mathrm{\frac{(\textbf{s}(\phi)\textbf{K}^{eff}_{e})^{2}}{\textbf{s}^{T}(\phi)\boldsymbol{\Lambda}^{(\theta)-1}\textbf{s}(\phi)} = \gamma_{usf}}
   \label{equation 6}
\end{equation}
This condition adopts the effective stress intensity factor at emission $\mathrm{\textbf{K}^{eff}_{e}}$ which is defined for the in-plane and anti-plane stress components acting on the slip plane given by :
\begin{equation}
\boldsymbol{\sigma}_{\theta \alpha} = \mathrm{\textbf{K}_{(\alpha) e}^{eff}}/\sqrt{2 \pi r },
\end{equation}
These are related to the stress intensity factor at the crack-tip by the relation 
\begin{equation}
\mathrm{\textbf{K}_{(\alpha) e}^{eff} = \textbf{F}_{\alpha \beta}(\theta) \textbf{K}_{(\beta) e}}
\end{equation}
Here, $\mathrm{\textbf{F}_{\alpha \beta}}$ contains the $\mathrm{\theta}$-dependence of $\boldsymbol{\sigma}_{\theta \alpha}$ for $\boldsymbol{K}_{(\beta) e}$. The $\mathrm{\textbf{F}_{\alpha \beta}}$ is obtained from the elastic stress field (\ref{Stroh formalism}). The latter part of the denominator in equation (\ref{equation 6}) relates the $\boldsymbol{\Lambda}^{-1}$ from the main crack to the slip plane in the slip direction, \textbf{Q} being a rotation matrix through angle $\mathrm{\theta}$ along $\mathrm{x_{3}}$. \par
\begin{equation}
\boldsymbol{\Lambda}^{(\theta)-1} = \boldsymbol{Q}(\theta) \boldsymbol{\Lambda}^{-1} \boldsymbol{Q}^{T}(\theta)
\end{equation}
For pure mode I loading, the equation for dislocation emission simplifies to the following expression :
\begin{equation}
    \mathrm{K_{Ie} = \frac{\sqrt{\gamma_{usf} \textbf{s}^{T}(\phi) \boldsymbol{\Lambda}^{(\theta)-1} \textbf{s}(\phi)}}{\textbf{F}_{12}(\theta) cos(\phi) + \textbf{F}_{32}(\theta) sin(\phi)  } }
\end{equation}
Based on the postulate by Eleanor Mak et al., 2021 \cite{MAK2021104389}, a material is intrinsically ductile if the stress intensity at emission is less than the stress intensity at cleavage in mode I. Hence, a ductility parameter D can be defined as follows :
\begin{equation}
    \mathrm{D = \frac{K_{Ie}}{K_{Ic}} = \chi  \, \Theta}
\end{equation}
where \par
    $\mathrm{\Theta} = \mathrm{\sqrt{\frac{\gamma_{usf}}{\gamma_{s}}}}$ \par
    $\mathrm{\chi} = \mathrm{\frac{\sqrt{\Lambda_{22} s^{T}(\phi) \Lambda^{(\theta)-1} s(\phi)}}{\sqrt{2}  \left [ F_{12}(\theta) cos(\phi) + F_{32}(\theta) sin(\phi)\right ]}}$ \par
\hfill \break
If $\mathrm{D<1}$ the material is ductile, brittle otherwise. We decompose the ductility parameter D into two separate components $\Theta$ and $\chi$. $\Theta$ is a function of $\mathrm{\gamma_{usf}}$ and $\mathrm{\gamma_{s}}$ and does not depend upon the elastic constants, hence being anisotropically independent. $\chi$, on the other hand, is a function of the distribution of the elastic constants in the elastic stiffness matrix (not just the absolute values) and it captures the effects of anisotropy as well as crack geometries. We do this bifurcation in order to better understand the effects of material anisotropy on the ductility of a system as compared to an ideal isotropic system.

\subsection{Crack configuration}
\label{crack cofiguration}
Multiple crack systems are possible in BCC metals. Intrinsic stacking faults are shown to exist in the crystal in multiple slip planes such as $\{110\}$, $\{112\}$ and $\{111\}$ \cite{doi:10.1080/14786436808227500}. Out of these, we have considered the 1/2$\langle 111 \rangle \{110\}$ slip system as it is known to be one of the most active slip systems in BCC crystals \cite{byron1967plastic}. Here, 1/2$\langle 111 \rangle$ is the direction of the Burger's vector and $\{110\}$ is the slip plane. For the crack planes, there is evidence that the \{100\} and \{110\} crystallographic planes are favored in BCC crystals \cite{TYSON1973621}. \par
As seen in Fig.~\ref{Crack png}, multiple crack systems are possible for failure in mode I for the \{110\} family of crack planes and \{110\} family of slip planes depending upon the angles $\mathrm{\theta}$ and $\mathrm{\phi}$. In this study, we have fixed the crack plane as the (1$\bar{1}$0) and all 12 possible 1/2$\langle 111 \rangle \{110\}$ slip systems were studied (table \ref{crack orientation table}). It should be noted that while for non-coplanar crack-slip systems there are unique values of $\mathrm{\theta}$ and $\mathrm{\phi}$, there are infinite possible values of $\mathrm{\phi}$ for coplanar systems ($\mathrm{\theta}$=0). In these cases, we have selected arbitrary geometries.  

\subsection{Density functional theory calculations}
\subsubsection{Generation of BCC HEA supercells}
In this study, the crystal structures for performing DFT calculations were generated using Atom, Molecule, Material Software Kit (ATOMSK) \cite{HIREL2015212}. Crystals containing pure BCC refractory elements tungsten (W), tantalum (Ta), chromium (Cr) and vanadium (V) as well as binary, ternary and quarternary HEAs with varying concentrations of these alloying elements were generated. For each composition, 10 random configurations were developed and the computed values were taken as the mean of these configurations to remove the effects of the local chemical environments as well as to improve the numerical stability of our DFT results. Table~\ref{Table elastic constants} consolidates the compositions studied in this work. From here onward, for the sake of succinctness and ease of reading, we will refer to the index numbers of the HEAs assigned in table~\ref{Table elastic constants} instead of the compositions themselves. \par

\begin{figure}
    \includegraphics[width=.22\textwidth]{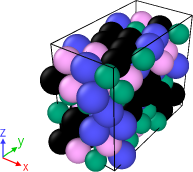}\hfill
    \includegraphics[width=.34\textwidth]{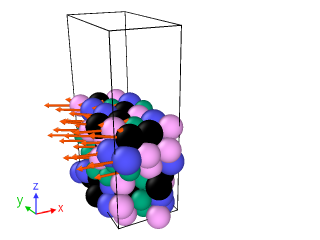}\hfill
    \caption{\textbf{a} A supercell bulk structure of an equimolar WTaCrV crystal visualized using OVITO \cite{ovito}. The black, blue, green and pink atoms represent W, Ta, Cr and V respectively. \textbf{b} An unstable stacking fault in the $\mathrm{[1 1 1]}$ direction in the $\mathrm{[1 \bar{1} 0]}$ plane induced in the bulk structure with free surfaces. The red arrows indicate a shift of distance of $\mathrm{a_0\sqrt{3}/4}$ in the $\mathrm{[1 1 1]}$ direction in the $\mathrm{[1 \bar{1} 0]}$ plane.}\label{ovito}
\end{figure}

Each supercell contains 108 atoms (see Fig. \ref{ovito}) and is in the shape of a rectangular prism. These supercells have orthogonal basis vectors $\mathrm{6[111], 2[\bar{1} \bar{1} 2] , 3[1 \bar{1} 0]}$ for the ease of calculations and visualization of the fracture and slip systems.
\subsubsection{Ground state energies of crystal structures}
This study uses the Vienna Ab initio Simulation Package (VASP) \cite{Hafner1997} to perform the DFT calculations. The structures were relaxed to the ground state at zero pressure (deformation of the supercell shape, volume and atomic positions allowed). We have used projector-augmented wave pseudopotentials
and exchange-correlation interactions described by the
generalized gradient approximation of the Perdew-Burke-Ernzerhof form. We employed a Methfessel-Paxton smearing
with width 0.2 eV and periodic boundary conditions in all directions
with a cutoff energy of planewaves of 500 eV. The Brillouin
zone was sampled with a Monkhorst-Pack scheme in a 3 $\times$ 3 $\times$ 3 k-point mesh for bulk samples and 3 $\times$ 3 $\times$ 1 k-point mesh for samples with free surfaces normal to the third axis. The convergence threshold for total energy and atomic force was $10^{-4}$ eV and $10^{-2}$ eV\AA$^{-1}$, respectively. We did not consider spin polarization in this study, as its effect in the results was negligible. 

\subsubsection{Free surface energy}
The free surface energies were calculated for the $\mathrm{(1 \bar{1} 0)}$ plane. Two free surfaces were created by adding a vacuum buffer of 10 $\mathrm{\AA}$ normal to the $\mathrm{(1 \bar{1} 0)}$ plane using ATOMSK in a fully periodic system (see Fig.~\ref{ovito}). This structure is then relaxed to the ground state by allowing the constituent atoms to move while the cell shape and volume remain fixed. Once the ground state energies of these structures are achieved, the energies of the free surfaces are calculated as follows:
\begin{equation}
    \mathrm{\gamma_{s}^{hkl} = \frac{E^{slab}_{hkl} - E^{bulk}}{2 A^{slab}}, }
\end{equation}
where $\mathrm{\gamma_{s}^{hkl}}$ is the free surface energy of the (hkl) plane, $\mathrm{E_{hkl}^{slab}}$ is the ground state energy of the system with free surfaces, $\mathrm{E_{bulk}}$ is the ground state energy of the bulk structure and $\mathrm{A^{slab}}$ is the area of the surface. As stated above, the mean of 10 random configurations is considered the material value for all the calculated properties henceforth. \par

\subsubsection{Unstable stacking fault energy}
In this work, for the sake of simplified calculations, we use the generalized stacking fault energy in the $[111]$ direction for a shift distance of half the burgers vector. This is an approximation to the more accurate and rigorous method of finding the maxima of a plot between the shift distance and the generalized stacking fault energy. There are benchmark calculations that show that the difference between the values obtained by the shift method and by curve interpolation is negligible, around 0.00062 J/$\mathrm{m^{2}}$, and hence can be used interchangeably \cite{screening_HU2021116800}. The $\mathrm{\gamma_{usf}}$ is calculated using the following equation:
\begin{equation}
    \mathrm{\gamma_{usf} = \frac{E_{usf}^{shift(slab)} - E^{slab}}{A^{slab}}.}
\end{equation}
Here, $\mathrm{\gamma_{usf}}$ is the unstable stacking fault energy, $\mathrm{E_{usf}^{shift(slab)}}$ is the ground state energy of the structure with free surface after shifting the top half atoms above the middle (1$\mathrm{\bar{1}}$0) plane by a distance equal to half the magnitude of the burgers vector in the $[111]$ direction, and $\mathrm{E^{slab}}$ is the ground state energy of the structure with free surface.

\par
The atomic positions are then relaxed only in the direction parallel to the plane normal, [1$\mathrm{\bar{1}}$0], while keeping the degrees of freedom in-plane constant and the cell shape and volume fixed. Free surfaces are also added to prevent the formation of two unstable stacking faults instead of one due to periodic boundary conditions. 
\subsubsection{Elastic constants}
\label{Elastic constants}
Elemental W, Ta, Cr and V are all known to possess a BCC crystal structure. For this work, we make use of the valence electron concentration (VEC) of the HEAs under investigation to predict their potential for forming stable BCC structures \cite{Guo_10.1063/1.3587228}. There is experimental evidence revealing that this class of HEAs is expected to form stable BCC crystals \cite{doi:10.1126/sciadv.aav2002}. Based on these findings, we apply BCC symmetries to all the relaxed bulk files. Due to the cubic nature of these structures, we have only three independent elastic constants, $\mathrm{C_{11},\, C_{12} \, and \, C_{44}}$ in the stiffness tensor. A theoretical summary of the framework used to calculate the elastic constants is discussed below.\par
For linear elastic materials, the generalized Hooke's law is given as:
\begin{equation}
    \boldsymbol{\sigma}_{i} = \textbf{C}_{ij}\boldsymbol{\epsilon}_{j}
\end{equation}
Here, $ \boldsymbol{\sigma}_{i}$ is the stress, $\mathrm{\textbf{C}_{ij}}$ the stiffness matrix and $\boldsymbol{\epsilon}_{j}$ the strain in the Voigt notation \cite{Ma_dziarz_2019}. The elastic energy for an applied strain is given by:
\begin{equation}
    \mathrm{\Delta E_{V}(\epsilon_{i}) = E_{V}(\epsilon_{i}) - E_{V_{0}}(0) = \frac{V_{0}}{{2}} \sum_{i,j = 1} ^{6} \textbf{C}_{ij} \boldsymbol{\epsilon}_{i} \boldsymbol{\epsilon}_{j}}
    \label{eq:Eelas}
\end{equation}
Here, $\mathrm{V_0}$ and V are the volume of the structure before and after the application of the strain, respectively, $\mathrm{E_{V_{0}}(0)}$ is the total energy of the structure before strain is induced, $\mathrm{E_{V}(\boldsymbol{\epsilon}_{i})}$ is the total energy after strain is applied and $\mathrm{\Delta E_{V}(\boldsymbol{\epsilon}_{i})}$ is the total elastic energy of the structure. Since we have three independent elastic constants, we obtain three equations by applying tri-axial normal strain $\mathrm{(\delta , \delta , \delta , 0,0,0)}$, bi-axial normal strain $\mathrm{(\delta , \delta ,0,0,0,0)}$, and tri-axial shear strain $\mathrm{(0,0,0, \delta , \delta , \delta)}$. We thus respectively obtain equations : 
\begin{equation}
    \mathrm{\Delta E = \frac{3 V_{0} (C_{11} + 2 C_{12}) \delta^{2}}{2}}
    \label{eq:EelasTri}
\end{equation}
\begin{equation}
    \mathrm{\Delta E = V_{0} (C_{11} + C_{12}) \delta^{2}}
    \label{eq:EelasBi}
\end{equation}
\begin{equation}
    \mathrm{\Delta E = \frac{3 V_{0} (C_{44}) \delta^{2}}{2}}
    \label{eq:EelasShear}
\end{equation}
Here, $\mathrm{\delta}$ is the magnitude of the applied deformation. In this work, $\mathrm{\delta} = [-0.015, -0.01, -0.005, 0, 0.005, 0.010, 0.015] $ to ensure small deformation and the viability of the linear elastic framework.\par
 The elastic constants were calculated using the VASPKIT \cite{VASPKIT} package. All the bulk structures consisting of perfect BCC crystal lattices were relaxed to their ground state. These structures are then subjected to the strains as discussed above and then relaxed keeping the cell volume and shape constant. Equation (\ref{eq:Eelas}) is used to calculate the strain energies of the structures associated with the respective deformations. We repeat this procedure for each of the 10 realizations for a given composition and compute the average of the elastic energy for each strain mode. Equations (\ref{eq:EelasTri}), (\ref{eq:EelasBi}) and (\ref{eq:EelasShear}) are then used to calculate $\mathrm{C_{11},\, C_{12} \, and \, C_{44}}$. \par
 Using these elastic constants, we also study the effects of the concentrations of the alloying elements on the anisotropy of the HEAs. Elemental W is experimentally verified to be a highly isotropic material \cite{tungsten_isotropic_10.1063/1.5044519,Elastic_Constants_PhysRev.130.1324}. Also, the isotropic nature of materials makes the Rice model, used in this work to predict the intrinsic ductility, significantly inexpensive in computational terms, opening up possibilities for much higher efficiencies and much broader compositional spaces due to the elimination of the need to calculate elastic constants. Since the class on HEAs pertaining to the scope of this study is W-based, it is important to understand the effects of alloying on the elastic anisotropy. In this work, as a parameter to quantify and measure the elastic anisotropy, we compute the Zener anisotropy ratio \cite{Zener_PhysRevLett.101.055504,tungsten_isotropic_10.1063/1.5044519} for each alloy system. An alloy with a cubic crystal structure is said to be isotropic if the ratio approaches unity (e.g. Zener anisotropy ratio for W = 1.15) \cite{Elastic_Constants_PhysRev.130.1324}. Zener anisotropic ratio $\mathrm{A_{z}}$ is given as follows:
 \begin{equation}
     \mathrm{A_{z} =\frac{2 C_{44}}{C_{11} - C_{12}}}
 \end{equation}

\subsubsection{Effects of anisotropy on ductility}
\label{effects of anisotropy}
The parameter $\mathrm{\chi}$ from the ductility metric D defined in Sec. \ref{Stress intensity factors and ductility metric D} is the term that accounts for the material anisotropy. It is independent of the absolute values of the elastic constants but is a function of the elastic anisotropy of the material. We use the values of Zener anisotropy ratios computed in Sec. \ref{Elastic constants} to quantify the anisotropy of the HEAs. The Irwin energy release due to elastic forces on the crack depends upon the crack orientation (Eq. (\ref{Equation 1})) and the relation is quantified by a relevant anisotropic elastic matrix defined in Sec. \ref{Stress intensity factors and ductility metric D}. \par

While the effects of the crack orientation on intrinsic ductility are encapsulated by the parameter $\boldsymbol{\Lambda}$, the anisotropy also affects the angular dependence of shear stress $\boldsymbol{\sigma}_{\theta \alpha}$ acting in the slip plane. This angular dependence is given by the terms $\mathrm{\textbf{F}_{\alpha \beta}}$ defined in Sec.\ref{Stress intensity factors and ductility metric D}. Data from previous research on BCC refractory HEAs (MoNbTiV) \cite{Ductile_and_brittle_crack-tip_response} shows that the values of $\mathrm{F_{\alpha \beta}}$ for alloys with moderate degree of anisotropy (Zener anisotropy ratio, $a_r=\frac{2C_{44}}{C_{11}-C_{22}}$, between 0.5 and 1.5) can be approximated to isotropic values. Hence, for ease of computation, we use the isotropic values for $\mathrm{\textbf{F}_{\alpha \beta}}$ given by $\mathrm{\textbf{F}_{12} = [sin(\theta/2) cos(\theta/2)^2]}$ and $\mathrm{\textbf{F}_{32} = [cos(\theta/2)]}$ \cite{Ductile_and_brittle_crack-tip_response,Bower2009-xj}. The values of Zener anisotropy ratio to justify this assumption are given in Fig.~\ref{zener anisotropy}, which indicates that the alloys formed are mostly isotropic in nature.

 \section{Phenomenological models for ductility prediction}\label{sec:phenomeno}
 The ductility metric discussed in Sec. \ref{Stress intensity factors and ductility metric D} is an analytical solution within the LEFM framework and is hence expected to be accurate. However, this model requires $\mathrm{\gamma_{s} , \gamma_{usf}}$ and the elastic constants of an alloy to estimate the ductility, which can get computationally expensive for HEAs since these values are configurational dependent. Hence, there is a need to investigate other ductility criteria that are more efficient and understand their performance against the analytical model. This section discusses a few phenomenological/approximate models for a more efficient ductility prediction. This is because these models require at least one or more DFT calculated parameters less than the Rice model. Due to the very nature of multi-principle element alloys (MPEAs), the compositional space can get quite huge and generally require large number of unique samples for testing \cite{HATLER2025101201}. This can make using the computationally intensive Rice model an inefficient way to study the ductility. The correlation of these phenomenological models with the analytical ductility parameter D is hence inspected. Models like these can be used as a computationally cheaper alternative to the Rice model for initial screening when the composition space is large to obtain a more manageable number of alloys that can then be analyzed using the more accurate (analytical) but computationally expensive Rice model. The following sections discuss some of the most widely used phenomenological as well as approximate models used to predict ductility of metallic alloys.\par

 \subsection{Surrogate ductility index $\mathrm{D_{surrogate}}$}
 \label{Dsur}
 This model proposed by Yong-Jie Hu et al. \cite{screening_HU2021116800} is an efficient approximation to the Rice criterion. This model maintains the essence of the analytical formulation by quantifying intrinsic ductility as the resistance to fracture failure. However, in order to reduce the computational expense, this approach uses $\mathrm{\gamma_{s}}$ and $\mathrm{\gamma_{usf}}$ as surrogate parameters to $\mathrm{K_{Ic}}$ and $\mathrm{K_{Ie}}$, respectively. The new surrogate ductility metric is hence defined as:
 \begin{equation}
     \mathrm{D_{surrogate} = \frac{\gamma_{s}}{\gamma_{usf}}}
 \end{equation}
 For higher values of $\mathrm{D_{surrogate}}$, the energy for crack propagation is expected to be higher than the energy for dislocation emission and hence the ductility of a metal is expected to increase. This model eliminates the requirement of calculating the elastic constants. However, other than being an approximation to the analytical solution, this model does not account for the material anisotropy or the crack orientation. 

 \subsection{LLD criteria}
 This is a phenomenological ductility metric proposed by Prashant Singh et al.\cite{SINGH2023119104}. It is hypothesized that ductility can be attributed to quantum-mechanical phenomena like the distortion of the crystal lattice in the local chemical environment and the chemical disorder during the formation of these alloys. This will increase the electronic band dispersion and cause disorder broadening. Local lattice distortions, which are caused by the mismatch in atomic sizes and elastic moduli, result in the formation of multiple new dislocation pathways. There is also a correlation between the electronegativity differences due to the complex chemical environments and the change in vacancy formation energies as well as the migration barriers within these HEA crystals. Local lattice distortions also result in localized strains in the crystal structure and impede dislocation motion. This framework attempts to use the correlation between the local lattice distortions and VEC to predict ductility. A dimensionless parameter of quantum-mechanical origin called LLD was defined as:
 \begin{equation}
     \mathrm{LLD = \Delta w_{VEC} \frac{\Delta \bar{\textbf{u}}_{x,y,z}}{\sqrt{(\bar{\textbf{u}}_{x,y,z})^2}}}
 \end{equation}
If the value of LLD is lower than 0.3, the material is likely to be ductile. $\mathrm{\Delta w_{VEC}}$ is the weighted VEC given by $\mathrm{ VEC_{HEA}^{bcc} - [VEC_{max}^{bcc} - VEC_{min}^{bcc}]}$ \cite{SINGH2023119104}. For the alloy to tend to form a BCC structure the range $\mathrm{[VEC_{max}^{bcc} - VEC_{min}^{bcc}]}$ is ($6-4 =2$). The parameters $\mathrm{\Delta \bar{\textbf{u}}_{x,y,z}}$ and $\mathrm{\sqrt{(\bar{\textbf{u}}_{x,y,z})^2}}$ are the average atomic displacements and the $\mathrm{L_{1,2}}$ norm of the average atomic displacements of the atoms during the relaxation of a perfect crystal structure to its ground state.

\subsection{Pugh ratio}
The Pugh ratio \cite{Pugh_1doi:10.1080/14786440808520496,pugh_2_THOMPSON2018100,pugh_3_PMID:33633140} is an empirical model based on the competition between plasticity and fracture. It is one of the most widely used ductility criterion for metals, mainly due to ease of use as it requires only the stiffness matrix, and the fact that it has successfully estimated ductility in some cases \cite{FENG2022110820}. This approach eliminates the need to compute free surface and unstable stacking fault energies. This phenomenological model is based on observations of direct scaling between the bulk modulus $K$ of a metal with the fracture stresses. The yield stress from the Orowan bowing equation is given by $\mathrm{\sigma_{Y} = Gb/\lambda}$. Pugh assumed that $\mathrm{\lambda}$ remains constant for metals, i.e., the effects of work hardening can be neglected. He related the Brinell hardness number (B.H.N.) to this by the expression $\mathrm{B.H.N. = Gb/c}$, where G is the shear modulus, b the Burger's vector and c a material constant. Pugh proposed that the strain state at the crack tip will cause the relevant elastic moduli to vary between the Young's modulus and $K$, and since these parameters scale together, chose $K$ for convenience. Furthermore, it was proposed that the fracture stress is proportional to $K$ and the lattice parameter a, i.e. $\mathrm{\sigma_{*} \propto Ka}$ \cite{pugh_2_THOMPSON2018100}. Assuming that the parameter b/ca remains constant for a crystal structure, we get the relation B.H.N./$\mathrm{\sigma_{*}}$ $\mathrm{\propto}$ G/K. This equation relates the ratio of the hardness to the fracture stress of a metal to the ratio of the shear modulus to the bulk modulus (Pugh ratio). For isotropic cubic crystals $K = (2\mathrm{C_{12} + C_{11}})/3$ and $G = \mathrm{(3C_{44}-C_{12}+C_{11})/5}$ \cite{pugh_3_PMID:33633140}. These expressions were used to formulate a relation that scales the ratio of G/K directly to the B.H.N. and inversely to the fracture stress. Based on these observations, the ductility criterion based on Pugh's ratio is given by:
\begin{equation}
    \mathrm{Pugh \: ratio = \frac{G}{K}}
\end{equation}
It is predicted that for lower values of Pugh's ratio the material will be softer (lower B.H.N.) with high fracture stresses (higher activation energy for fracture) and hence more ductile. Pugh did not propose a critical value of G/K for the ductile to brittle transition of metals, but further investigation suggested the value to be around 0.57 to 0.6 \cite{pugh_2_THOMPSON2018100}.

\subsection{Statistical analysis}
Each statistical correlation studied in this work is done using the Pearson correlation coefficient given by the formula
\begin{equation}
    \mathrm{R=\frac{(x_{i}-\bar{x}) \, (y_{i}-\bar{y})}{\sqrt{\Sigma(x_{i}-\bar{x})^{2} \, \Sigma(y_{i}-\bar{y})^{2}}}},
\end{equation}
where R is the Pearson correlation coefficient, $\mathrm{x_{i}}$ and $\mathrm{y_{i}}$ are the values of the x and y variables in the sample and $\mathrm{\bar{x}}$ and $\mathrm{\bar{y}}$ are the mean values of x and y variables, respectively. Values of R closer to 1 and -1 indicate strong positive and negative correlations, respectively, and values close to 0 indicate weak correlations. \par
The p values associated with the correlation coefficient relate to the statistical significance of the correlation. They represent the probability of observing a correlation erroneously without any real correlation between the variables. Lower values of p ($\mathrm{\textless}$0.05) indicate that the observed correlations are statistically significant and less likely to have occurred by chance.

\section{Results and Discussions}

 \begin{figure}
    \centering
    \includegraphics[width=3.4in]{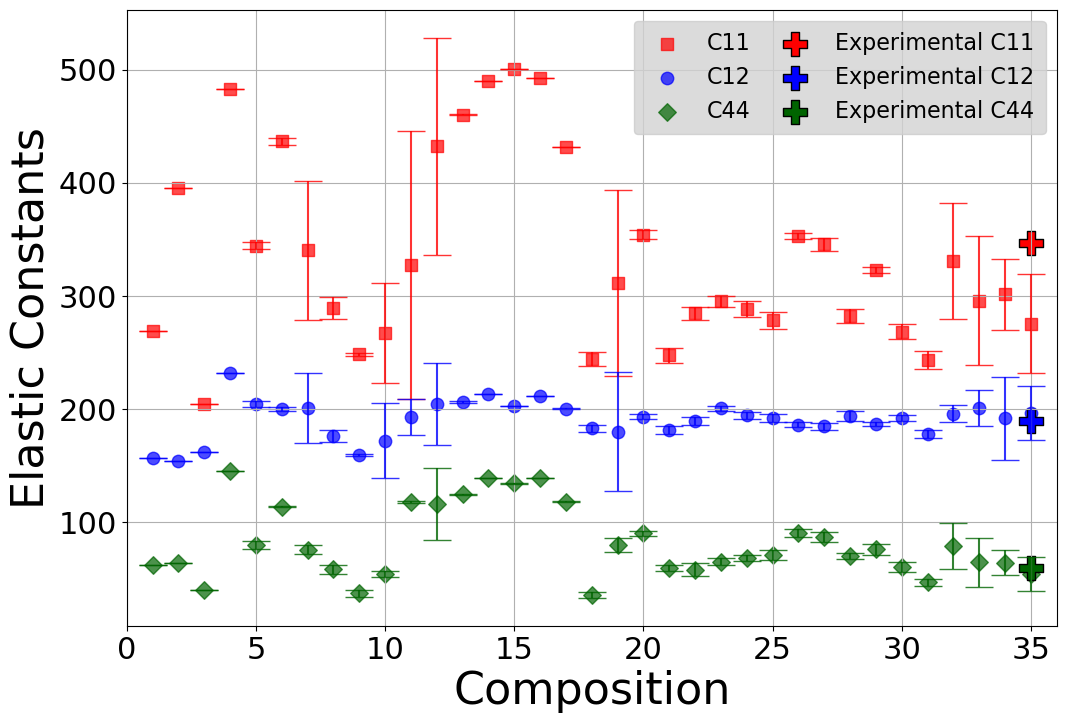}
    \caption{The theoretical elastic constant values for the HEAs and the experimental values for $\mathrm{W_{38}Ta_{36}Cr_{15}V_{11}}$. The error bars represent the variation in the computed values for the 10 different configurations as discussed in section \ref{Elastic constants}. }\label{Elastic experimenta}
\end{figure}

\subsection{Elastic anisotropy}
\begin{figure}
    \includegraphics[width=3.5in]{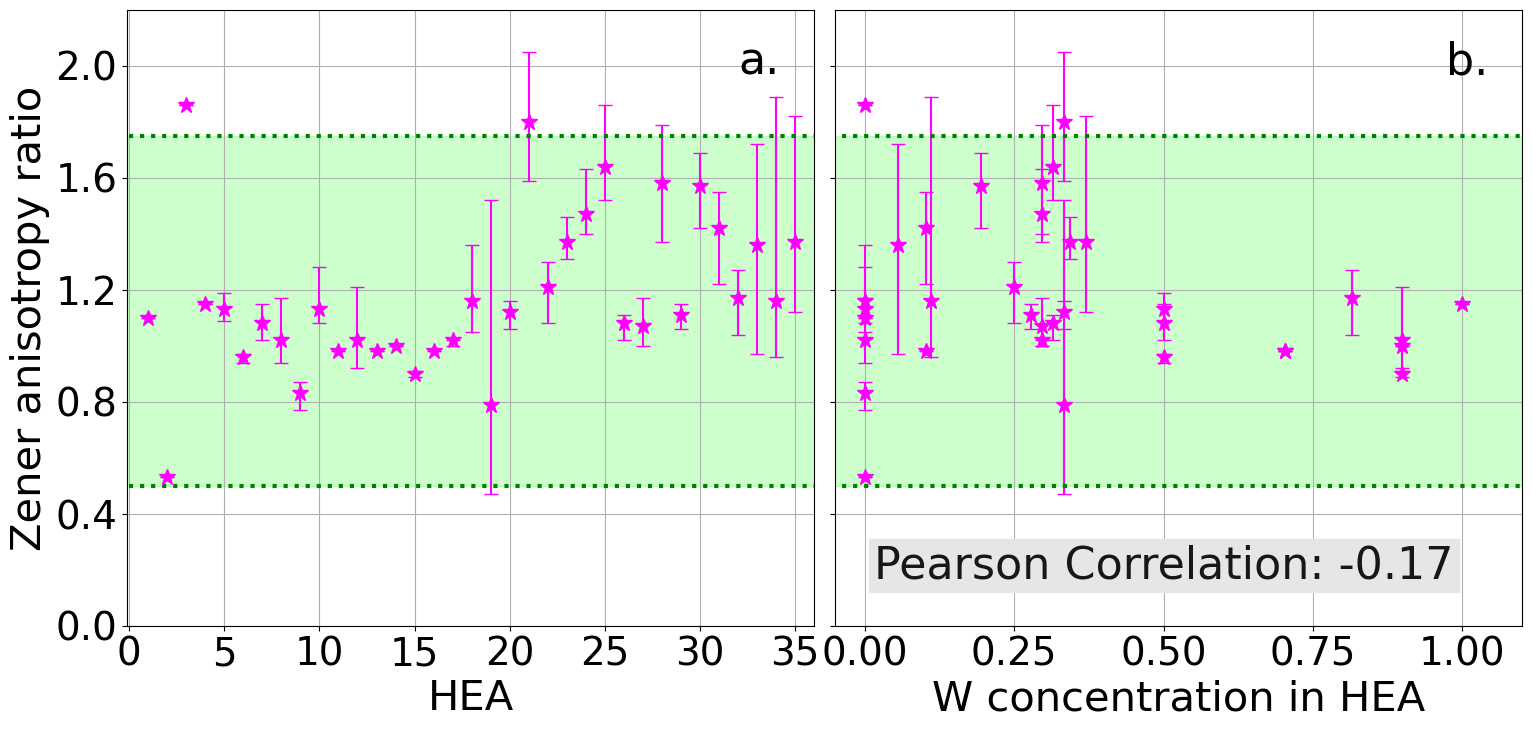}
    \caption{\textbf{(a.)} The computed values of Zener anisotropy ratio for the WTaCrV HEAs. \textbf{(b.)} The variation of the Zener anisotropy ratio with the concentration of W in the HEA. The error bars represent the variation in the computed values for the 10 different configurations as discussed in section \ref{Elastic constants}. }
\label{zener anisotropy}
\end{figure}

Table \ref{Table elastic constants} contains the DFT computed values of the three independent elastic constants, $\mathrm{C_{11},\, C_{12} \, and \, C_{44}}$ for each WTaCrV HEA studied in this work. These values are compared to experimental data for $\mathrm{W_{38}}$$\mathrm{Ta_{36}}$$\mathrm{Cr_{15}}$$\mathrm{V_{11}}$ ($\mathrm{C_{11} = 347 ~GPa}$, $\mathrm{C_{12} = 189 ~GPa}$, $\mathrm{C_{44} = 59 ~GPa}$) which show small errors and hence validate the robustness of our DFT model (Fig. \ref{Elastic experimenta}). Tungsten is known to be isotropic \cite{tungsten_isotropic_10.1063/1.5044519,Elastic_Constants_PhysRev.130.1324}, but the properties of these W-based HEAs are mainly unknown. Our findings show that this class of HEAs is isotropic in nature (see Sec.~\ref{effects of anisotropy}). Figure \ref{zener anisotropy} (b) shows that these HEAs remain fairly isotropic regardless of the concentration of W in the HEA and that there is a negligible correlation between alloying and anisotropy (Pearson correlation = -0.17). \par
The values of the parameter $\chi$ from Rice model are also calculated for all 12 crack orientations and tabulated in table \ref{table chi}. As expected from their isotropic nature, our results show that the values of the anisotropic ductility parameter $\chi$, which encapsulates the effects of anisotropy, do not vary much for different compositions. We hence treat the values of $\chi$ as constants that depend only on the crack orientation and not on the HEA composition. \par

\subsection{Ductility predictions using the Rice model}
A ductility assessment of 33 WTaCrV HEAs for varying concentrations was carried out using the Rice model for the $\{110\}$ crack planes and $\langle 111\rangle \{110\}$ slip system for 12 different crack orientations. Table \ref{crack orientation table} shows the angles $\mathrm{\theta}$ and $\mathrm{\phi}$ that quantify the studied crack geometries along with the crack planes and slip systems, as defined in Sec. \ref{Geometry}. The surface energies and the unstable stacking fault energies remain the same for the family of \{110\} crack planes and $\langle 111\rangle \{110\}$ slip systems. This results in the $\mathrm{\Theta}$ defined in Sec. \ref{Stress intensity factors and ductility metric D} to remain a constant value independent of the crack orientation. The change in intrinsic ductility for an HEA with respect to different crack orientations would thus be a function of the parameter $\mathrm{\chi}$ (see Sec. \ref{Stress intensity factors and ductility metric D}).  

\begin{figure*}
  \includegraphics[width=18cm]{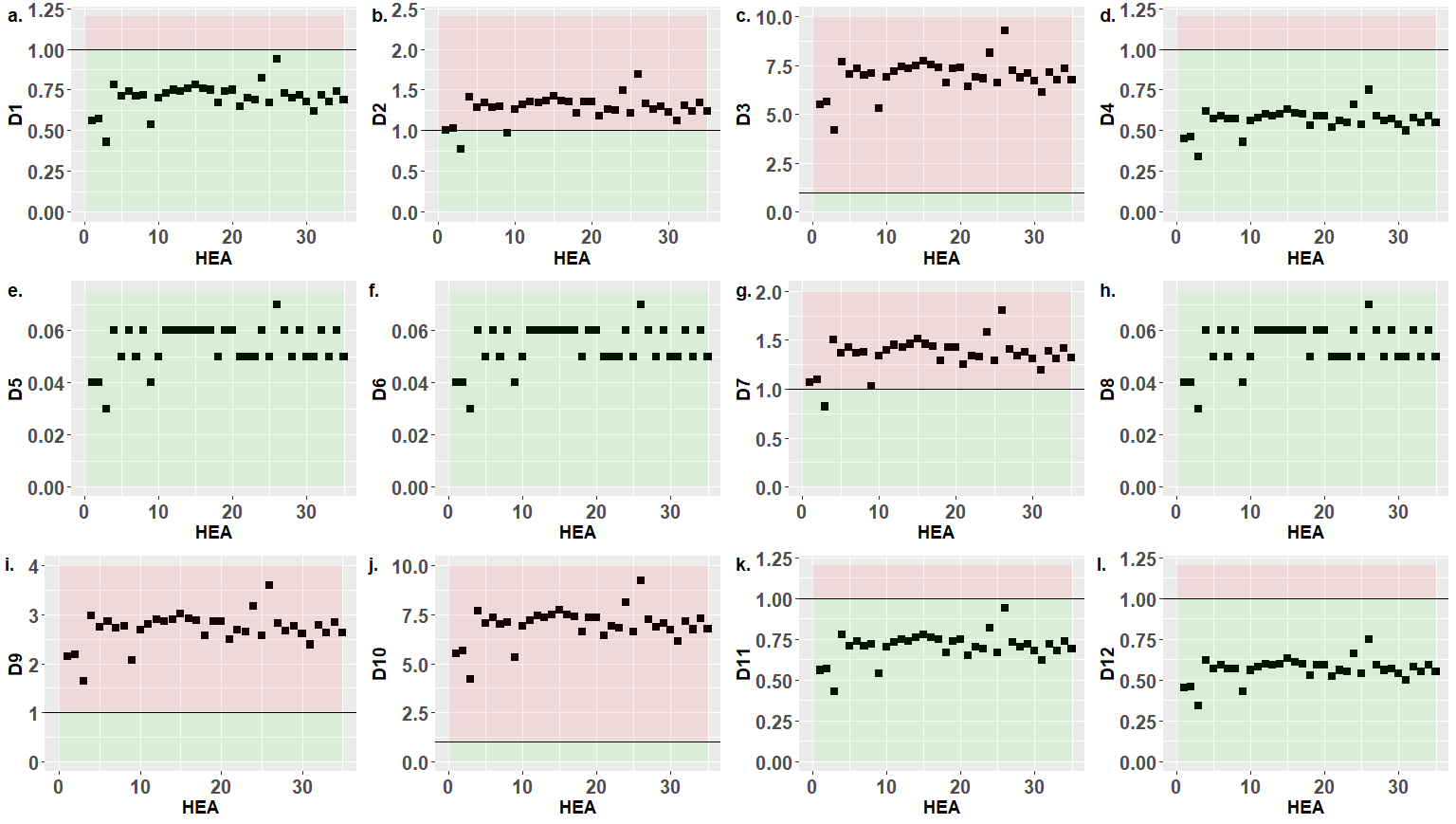}
  \caption{$\mathrm{\textbf{D parameter}}$ \textbf{(a)}-\textbf{(l)} The ductility parameter D computed using the Rice model for each HEA for the studied crack systems. The solid horizontal lines represent the threshold (D=1) imposed by the Rice model for an alloy to be ductile. The plot shaded in green predicts the alloy to be ductile and the red region predicts the alloy to be brittle. Figures were generated using R studio \cite{Rstudio} }
  \label{Rice png}
\end{figure*}

The results for the intrinsic ductility of each HEA for all 12 crack geometries are tabulated in table \ref{Phenomenological table}. It is observed that 7 out of the 12 crack systems studied predict that the alloys are in the ductile range for the D values defined in Sec. \ref{Stress intensity factors and ductility metric D}. The data suggest that these alloys show potential to be intrinsically ductile. The results are tabulated in table \ref{Phenomenological table} and Fig. \ref{Rice png}. The solid horizontal lines in the figures indicate the threshold of D to exhibit intrinsic ductility as proposed by the Rice model. All the HEAs below the line (green region of the plot) are predicted to be ductile for these crack systems. \par

\begin{figure}
  \includegraphics[width=9cm]{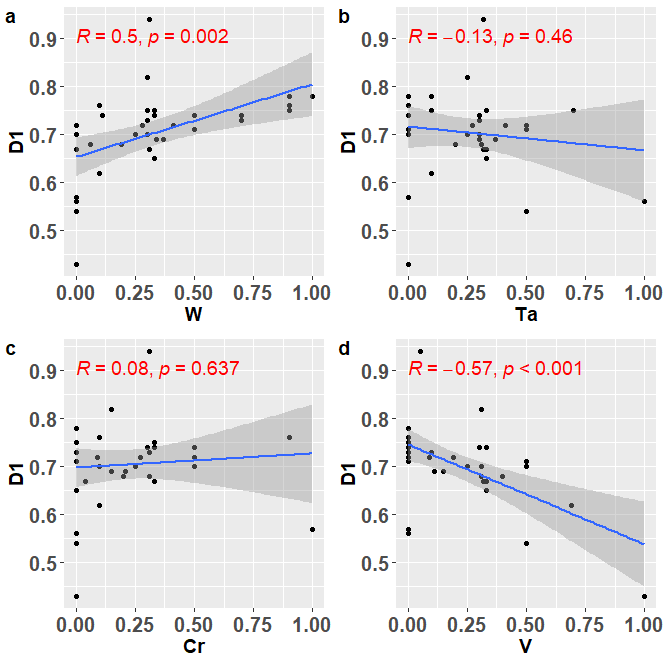}
  \caption{\textbf{Correlations of the D parameter for the crack systems 1 vs the concentrations of alloying elements.}\textbf{(a)} D1 values vs W concentration in the HEA. \textbf{(b)} D1 values vs Ta concentration in the HEA. \textbf{(c)} D1 values vs Cr concentration in the HEA. \textbf{(d)} D1 values vs V concentration in the HEA.  The values for all crack systems (D1-D12) can be found in table \ref{Phenomenological table}.}
  \label{D element png}
\end{figure}

 The correlations between the atomic concentrations of the alloys and the ductility parameters were also studied. The correlation between crack system 1 (D1 values in table \ref{Phenomenological table}) and the concentrations of the alloying elements are shown in Fig. \ref{D element png}. Our findings indicate that the concentration of W in the HEA tends to make the alloy more brittle by increasing the value of the ductility parameter D with R=0.5 (p=0.002), indicating a strong negative correlation with ductility (ductility increases with lower values of D). An increase in V concentration, on the other hand, makes the alloy more ductile with R=-0.57 (p\textless0.001) indicating strong positive correlation with ductility. Ta and Cr concentrations do not seem to correlate strongly to the intrinsic ductility for these isotropic HEAs and hence are found not to actively influence ductility. The correlations between D values and W concentrations lie within the range R=[0.5,0.51], for Ta concentrations in the range R=[-0.14,-0.12], for Cr concentrations in the range R=[0.08,0.1] and for V concentrations in the range R=[-0.6,-0.57], for all 12 slip systems. This means that the correlations between the D values and the elemental concentrations in the HEAs maintain the same trend for all 12 slip systems as in Fig. \ref{D element png} and that we can translate the results shown for crack orientation D1 to all 12 slip systems. The exact values for all crack orientations are shown in the supplementary text. Figure \ref{D Phenomenological png}(d) also studies the values of D1 with respect to $\mathrm{\Theta}$. $\mathrm{\Theta}$ represents the ductility metric without considering the effects of anisotropy. This means that an isotropic system will have a constant value of $\mathrm{\chi}$ and only $\mathrm{\Theta}$ will be an intrinsic variable while computing the D parameter. A strong positive correlation of R=1 further strengthens the fact that the effects of anisotropy in these HEAs are negligible, and hence they can be considered fully isotropic in nature. \par

\begin{table}[htbp]
    \centering
    \begin{tabular}{|c|c|c|c|c|c|c|}
    \toprule
   Index & HEA & \thead{$\mathrm{\gamma_{s}}$ \\ ($\mathrm{J/m^{2}}$)} & \thead{$\mathrm{\gamma_{usf}}$ \\ ($\mathrm{J/m^{2}}$)} &  \thead{$\mathrm{C_{11}}$ \\ (GPa)} & \thead{$\mathrm{C_{12}}$ \\ (GPa)} & \thead{$\mathrm{C_{44}}$ \\ (GPa)}    \\
    \midrule
 1 & $\mathrm{Ta}$  & 2.35  & 0.68  & 269   & 157   & 62    \\
  2  &  $\mathrm{Cr}$  & 3.27  & 1.02  & 395   & 154   & 64     \\
   3  & $\mathrm{V}$  & 3.48  & 0.6   & 204   & 162   & 39     \\
    4  &    $\mathrm{W}$  & 3.27  & 1.82  & 483   & 231   & 145    \\
     5  &   $\mathrm{W_{50}}$$\mathrm{Ta_{50}}$  & 2.84  & 1.33  & 344   & 204   & 80   \\
      6  &  $\mathrm{W_{50}}$$\mathrm{Cr_{50}}$ & 3.08  & 1.57  & 437   & 200   & 113  \\
       7  & $\mathrm{W_{50}}$$\mathrm{V_{50}}$  & 2.9   & 1.35  & 340   & 201   & 75  \\
       8 &  $\mathrm{Ta_{50}}$$\mathrm{Cr_{50}}$  & 2.57  & 1.18  & 289   & 176   & 58  \\
       9 &  $\mathrm{Ta_{50}}$$\mathrm{V_{50}}$  & 2.28  & 0.71  & 248   & 159   & 37  \\
        10 & $\mathrm{Cr_{50}}$$\mathrm{V_{50}}$  & 2.84  & 1.28  & 267   & 172   & 54  \\
         11& $\mathrm{W_{70}}$$\mathrm{Ta_{30}}$  & 2.93  & 1.43  & 327   & 193   & 118   \\
         12& $\mathrm{W_{90}}$$\mathrm{Ta_{10}}$  & 3.04  & 1.6   & 432   & 204   & 116  \\
         13& $\mathrm{W_{70}}$$\mathrm{Cr_{30}}$  & 3.14  & 1.6   & 460   & 206   & 124  \\
         14& $\mathrm{W_{90}}$$\mathrm{Cr_{10}}$  & 3.21  & 1.7   & 490   & 213   & 138  \\
        15 & $\mathrm{W_{90}}$$\mathrm{Ta_{10}}$  & 3.26  & 1.86  & 500   & 203   & 134  \\
        16 & $\mathrm{W_{10}}$$\mathrm{Cr_{90}}$  & 3.23  & 1.73  & 493   & 212   & 138  \\
        17 & $\mathrm{W_{30}}$$\mathrm{Ta_{70}}$  & 2.99  & 1.54  & 432   & 200   & 118  \\
       18  & $\mathrm{Ta_{33}}$$\mathrm{Cr_{33}}$$\mathrm{V_{33}}$  & 2.47  & 1.02  & 244   & 183   & 36    \\
       19  & $\mathrm{W_{33}}$$\mathrm{Cr_{33}}$$\mathrm{V_{33}}$  & 2.94  & 1.51  & 311   & 180   & 79     \\
       20  & $\mathrm{W_{33}}$$\mathrm{Ta_{33}}$$\mathrm{Cr_{33}}$ & 2.82  & 1.45  & 354   & 193   & 90     \\
        21 & $\mathrm{W_{33}}$$\mathrm{Ta_{33}}$$\mathrm{V_{33}}$ & 2.65  & 1.04  & 247   & 181   & 59    \\
       22  & $\mathrm{W_{25}}$$\mathrm{Ta_{25}}$$\mathrm{Cr_{25}}$$\mathrm{V_{25}}$  & 2.82  & 1.28  & 284   & 189   & 58   \\
      23   & $\mathrm{W_{34}}$$\mathrm{Ta_{30}}$$\mathrm{Cr_{21}}$$\mathrm{V_{15}}$  & 2.87  & 1.27  & 295   & 200   & 65   \\
       24  & $\mathrm{W_{30}}$$\mathrm{Ta_{25}}$$\mathrm{Cr_{15}}$$\mathrm{V_{31}}$  & 2.35  & 1.47  & 288   & 194   & 68     \\
      25   & $\mathrm{W_{31}}$$\mathrm{Ta_{32}}$$\mathrm{Cr_{5}}$$\mathrm{V_{32}}$  & 2.63  & 1.1   & 278   & 192   & 71    \\
       26  & $\mathrm{W_{31}}$$\mathrm{Ta_{32}}$$\mathrm{Cr_{31}}$$\mathrm{V_{6}}$  & 2.25  & 1.82  & 353   & 186   & 90    \\
       27  & $\mathrm{W_{30}}$$\mathrm{Ta_{30}}$$\mathrm{Cr_{31}}$$\mathrm{V_{9}}$  & 2.78  & 1.4   & 345   & 185   & 86  \\
       28  & $\mathrm{W_{30}}$$\mathrm{Ta_{30}}$$\mathrm{Cr_{9}}$$\mathrm{V_{31}}$ & 2.63  & 1.18  & 282   & 193   & 70    \\
       29  & $\mathrm{W_{28}}$$\mathrm{Ta_{27}}$$\mathrm{Cr_{27}}$$\mathrm{V_{18}}$  & 2.74  & 1.31  & 323   & 186   & 76    \\
      30   & $\mathrm{W_{19}}$$\mathrm{Ta_{20}}$$\mathrm{Cr_{20}}$$\mathrm{V_{41}}$  & 2.63  & 1.12  & 268   & 192   & 60    \\
        31 & $\mathrm{W_{10}}$$\mathrm{Ta_{10}}$$\mathrm{Cr_{10}}$$\mathrm{V_{70}}$  & 2.56  & 0.91  & 243   & 178   & 47    \\
       32  & $\mathrm{W_{41}}$$\mathrm{Ta_{41}}$$\mathrm{Cr_{9}}$$\mathrm{V_{9}}$  & 2.77  & 1.34  & 331   & 196   & 79     \\
       33  & $\mathrm{W_{7}}$$\mathrm{Ta_{31}}$$\mathrm{Cr_{31}}$$\mathrm{V_{31}}$  & 3.07  & 1.33  & 296   & 201   & 64    \\
       34  &  $\mathrm{W_{10}}$$\mathrm{Ta_{30}}$$\mathrm{Cr_{30}}$$\mathrm{V_{30}}$  & 2.7   & 1.36  & 301   & 191   & 64    \\
        35  & $\mathrm{W_{38}}$$\mathrm{Ta_{36}}$$\mathrm{Cr_{15}}$$\mathrm{V_{11}}$  & 2.92  & 1.27  & 294   & 210   & 62   \\
    \bottomrule
    \end{tabular}
 \captionof{table}{The values of $\mathrm{\gamma_{s}}$, $\mathrm{\gamma_{USF}}$ and elastic constants $\mathrm{C_{11}}$, $\mathrm{C_{12}}$ and $\mathrm{C_{44}}$ for each HEA.} 
 \label{Table elastic constants}
\end{table}

\begin{center}

    \begin{table*}[ht]
    \centering
    \begin{tabular}{|c|c|c|c|c|c|c|c|}%
    \toprule
    \thead{\normalsize System} & 
    \thead{ \normalsize Crack \\ \normalsize orientation \\ $(\mathrm{x_{1}}-\mathrm{x_{2}}-\mathrm{x_{3}})$}  &
    \thead{\normalsize Slip \\ \normalsize system} &
    \thead{\normalsize $\mathrm{\theta (^{o})}$} & 
    \thead{\normalsize $\mathrm{\phi (^{o})}$} &
    \thead{\normalsize $\mathrm{\textbf{s}(\phi)}$} & 
    \thead{\normalsize $\mathrm{\textbf{F}_{12}^{isotropic}}$ } & 
    \thead{\normalsize $\mathrm{\textbf{F}_{32}^{isotropic}}$} \\
    \midrule
     1 & $\mathrm{[\bar{1}1\bar{2}] -[110] - [1\bar{1} \bar{1}]}$ & $\mathrm{1/2[1 1 \bar{1}] (101)}$ & 60 & 19.47 &  $\mathrm{[0.94,0,0.33]^{T}}$ & 0.37 & 0.87 \\
     \midrule
     2 & $\mathrm{[1\bar{1}\bar{2}]-[110]-[1\bar{1}1]}$ & $\mathrm{1/2[1 \bar{1} 1] (011)}$ & 60 & 0.00 &  $\mathrm{[1,0,0]^{T}}$ & 0.37 & 0.87 \\
     \midrule
     3 & $\mathrm{[1\bar{1}\bar{2}]-[110]-[1\bar{1}1]}$ & 1/2$\mathrm{[11\bar{1}](011)}$ & 60  & -19.47 & $\mathrm{[0.94, 0, -0.33]^{T}}$ & 0.37 & 0.87 \\
     \midrule
     4 & $\mathrm{[\bar{1}10]-[110]-[00\bar{1}]}$ & 1/2$\mathrm{[11\bar{1}](1\bar{1}0)}$ & 90  & 35.26 & $\mathrm{[0.82, 0, 0.58]^{T}}$ & 0.35 & 0.71 \\
     \midrule
     5 & $\mathrm{[1\bar{1}2]-[110]-[1\bar{1}\bar{1}]}$ & 1/2$\mathrm{[1\bar{1}\bar{1}](01\bar{1})}$ & 60  & 90 & $\mathrm{[0, 0, 1]^{T}}$ & 0.37 & 0.87 \\
     \midrule
     6 & $\mathrm{[\bar{1}1\bar{2}]-[110]-[1\bar{1}\bar{1}]}$ & 1/2$\mathrm{[1\bar{1}\bar{1}](101)}$ & 60  & 90 & $\mathrm{[0, 0, 1]^{T}}$ & 0.37 & 0.87 \\
     \midrule
     7 & $\mathrm{[\bar{5}52]-[110]-[\bar{1}1\bar{5}]}$ & 1/2$\mathrm{[1\bar{1}\bar{1}](110)}$ & 0  & 19.75 & $\mathrm{[0.94, 0, 0.33]^{T}}$ & 0 & 1 \\
     \midrule
     8 & $\mathrm{[\bar{1}12]-[110]-[1\bar{1}1]}$ & 1/2$\mathrm{[1\bar{1}1](10\bar{1})}$ & 60  & 90 & $\mathrm{[0, 0, 1]^{T}}$ & 0.37 & 0.87 \\
     \midrule
     9 & $\mathrm{[3\bar{3}2]-[110]-[\bar{1}13]}$ & 1/2$\mathrm{[1\bar{1}1](110)}$ & 0  & 10.02 & $\mathrm{[0.98, 0, 0.17]^{T}}$ & 0 & 1 \\
     \midrule
     10 & $\mathrm{[1\bar{1}2]-[110]-[\bar{1}11]}$ & 1/2$\mathrm{[111](01\bar{1})}$ & 60  & 19.47 & $\mathrm{[0.94, 0, -0.33]^{T}}$ & 0.37 & 0.87 \\
     \midrule
     11 & $\mathrm{[\bar{1}12]-[110]-[\bar{1}1\bar{1}]}$ & 1/2$\mathrm{[111](10\bar{1})}$ & 60  & -19.47 & $\mathrm{[0.94, 0, 0.33]^{T}}$ & 0.37 & 0.87 \\
     \midrule
     12 & $\mathrm{[\bar{1}10]-[110]-[001]}$ & 1/2$\mathrm{[111](1\bar{1}0)}$ & 90  & 35.26 & $\mathrm{[0.82, 0, 0.58]}$ & 0.35 & 0.71 \\

     \bottomrule
     \end{tabular}%
     \captionof{table}{Crack systems with the crack orientations ($\mathrm{x_{1}}$, $\mathrm{x_{2}}$, $\mathrm{x_{3}}$ represent the crack propagation direction, direction normal to the crack plane and the direction of the crack front, respectively), slip planes along with the angles $\mathrm{\theta (^{o})}$ and $\mathrm{\phi (^ {o})}$ as defined in \ref{Geometry}. For isotropic materials, $F_{12}=cos^{2}(\theta / 2) sin(\theta / 2)$ and $F_{32} = cos(\theta /2)$ \cite{RICE1992239}. $\mathrm{\textbf{s}(\phi)}$ is given in the rotated basis as described in Sec. \ref{Geometry}.}\label{crack orientation table}
     \end{table*}
     
\end{center}

\begin{table*}[htbp]
  \centering
    \begin{tabular}{|c|c|c|c|c|c|c|c|c|c|c|c|}
    \toprule
    $\mathrm{\chi_{1}}$ & $\mathrm{\chi_{2}}$ & $\mathrm{\chi_{3}}$ & $\mathrm{\chi_{4}}$ & $\mathrm{\chi_{5}}$ & $\mathrm{\chi_{6}}$ & $\mathrm{\chi_{7}}$ & $\mathrm{\chi_{8}}$ & $\mathrm{\chi_{9}}$ & $\mathrm{\chi_{10}}$ & $\mathrm{\chi_{11}}$ & $\mathrm{\chi_{12}}$ \\
    \midrule
    1.04  & 1.88  & 10.28 & 0.83  & 0.08  & 0.08  & 2.00  & 0.08  & 4     & 10.28 & 1.04  & 0.83 \\
    \bottomrule
    \end{tabular}%
    \caption{The values of the anisotropic part of the ductility parameter $\mathrm{\chi}$ for all the 12 crack orientations studied in this work}
  \label{table chi}%
\end{table*}%

\subsection{Ductility predicted using phenomenological models}
\begin{figure}
  \includegraphics[width=9cm]{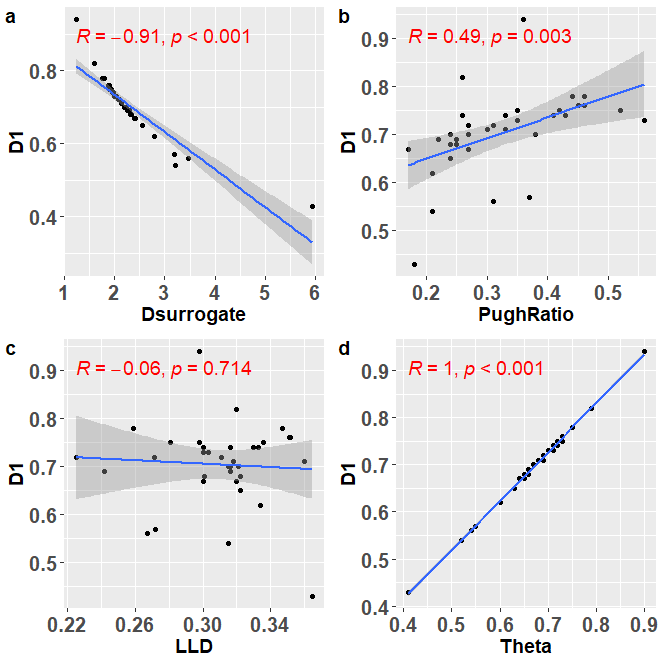}
  \caption{\textbf{Correlations of the D parameter for the crack systems 1 vs the phenomenological models.} 
  \textbf{(a)} D1 values vs $\mathrm{D_{surrogate}}$ values. \textbf{(b)} D1 values vs Pugh ratio values. \textbf{(c)} D1 values vs LLD values. \textbf{(d)} D1 values vs $\mathrm{\Theta}$ values.  The values for all crack systems (D1-D12) can be found in table \ref{Phenomenological table}.}
  \label{D Phenomenological png}
\end{figure}

All the alloys that are studied using the Rice model were also studied using the empirical approaches described above. The results obtained using these models are listed in table \ref{Phenomenological table}. These phenomenological models are analyzed against the analytical solutions obtained from the Rice model. Figure \ref{D Phenomenological png} shows the correlation between the D value for crack orientation 1 against the phenomenological models. Figure $\mathrm{D_{surrogate}}$ shows that both the $\mathrm{D_{surrogate}}$ and Pugh ratio have good correlations with the D1 values (R=-0.91 and R=0.49 respectively). On the other hand, the LLD values have a weak correlation (R=-0.06) with D1. The correlations of the other crack orientations (D1 to D12) with the $D_{surrogate}$ model has correlations in the range R=[-0.92,-0.87], for Pugh ratio R=[0.49,0.56] and for LLD R=[-0.1,-0.06]. This means that the correlations follow the trends seen with D1 and the phenomenological models as shown in Fig. \ref{D Phenomenological png}. Hence, we can translate and use the observed correlation trends for D1 to all crack orientations (D1 to D12). The exact results are provided in the supplementary text. Our findings suggest that the $\mathrm{D_{surrogate}}$ parameter defined in \ref{Dsur} can be an efficient way to screen the large compositional space of HEAs, as long as they are isotropic. The Pugh ratio is also found to be an effective screening method when computing the elastic constants is more desirable or easier or if the values of the elastic constants are already known. The last phenomenological model discussed in this work is the LLD model, which tries to correlate the distortions in the crystal lattice and the local chemical environments with the intrinsic ductility of the alloys. Our results show that this metric is not accurate to be used as ductility estimator for this class of WTaCrV alloys accounting to the insignificant correlations with the Rice model.

\begin{center}
    \begin{table*}
    \begin{tabular}{|c|c|c|c|c|c|c|c|c|c|c|c|c|c|c|c|c|}
    \hline  
     HEA & $\mathrm{\Theta}$ & $\mathrm{D_{1}}$ & $\mathrm{D_{2}}$  & $\mathrm{D_{3}}$  & $\mathrm{D_{4}}$  &$\mathrm{D_{5}}$ & $\mathrm{D_{6}}$   & $\mathrm{D_{7}}$  & $\mathrm{D_{8}}$  & $\mathrm{D_{9}}$   & $\mathrm{D_{10}}$  & $\mathrm{D_{11}}$  & $\mathrm{D_{12}}$  &  $\mathrm{D_{surrogate}}$ & $\mathrm{LLD}$  & \thead{Pugh \\ ratio} \\
     \hline
    1     & 0.54  & 0.56  & 1.01  & 5.51  & 0.45  & 0.04  & 0.04  & 1.07  & 0.04  & 2.15  & 5.51  & 0.56  & 0.45  & 3.46  & 0.27  & 0.31 \\
    \hline
    2     & 0.56  & 0.57  & 1.03  & 5.64  & 0.46  & 0.04  & 0.04  & 1.10  & 0.04  & 2.19  & 5.64  & 0.57  & 0.46  & 3.21  & 0.27  & 0.37 \\
    \hline
    3     & 0.42  & 0.43  & 0.77  & 4.21  & 0.34  & 0.03  & 0.03  & 0.82  & 0.03  & 1.64  & 4.21  & 0.43  & 0.34  & 5.80  & 0.36  & 0.18 \\
    \hline
    4     & 0.75  & 0.78  & 1.41  & 7.69  & 0.62  & 0.06  & 0.06  & 1.50  & 0.06  & 2.99  & 7.69  & 0.78  & 0.62  & 1.80  & 0.26  & 0.44 \\
    \hline
    5     & 0.68  & 0.71  & 1.29  & 7.05  & 0.57  & 0.05  & 0.05  & 1.37  & 0.05  & 2.74  & 7.05  & 0.71  & 0.57  & 2.14  & 0.32  & 0.30 \\
    \hline
    6     & 0.71  & 0.74  & 1.34  & 7.35  & 0.59  & 0.06  & 0.06  & 1.43  & 0.06  & 2.86  & 7.35  & 0.74  & 0.59  & 1.96  & 0.32  & 0.41 \\
    \hline
    7     & 0.68  & 0.71  & 1.28  & 7.02  & 0.57  & 0.05  & 0.05  & 1.37  & 0.05  & 2.73  & 7.02  & 0.71  & 0.57  & 2.15  & 0.36  & 0.33 \\
    \hline
    8     & 0.68  & 0.72  & 1.30  & 7.11  & 0.57  & 0.06  & 0.06  & 1.38  & 0.06  & 2.77  & 7.11  & 0.72  & 0.57  & 2.18  & 0.27  & 0.27 \\
    \hline
    9     & 0.56  & 0.54  & 0.97  & 5.32  & 0.43  & 0.04  & 0.04  & 1.03  & 0.04  & 2.07  & 5.32  & 0.54  & 0.43  & 3.21  & 0.32  & 0.21 \\
    \hline
    10    & 0.67  & 0.70  & 1.26  & 6.91  & 0.56  & 0.05  & 0.05  & 1.34  & 0.05  & 2.69  & 6.91  & 0.70  & 0.56  & 2.22  & 0.32  & 0.38 \\
    \hline
    11    & 0.70  & 0.73  & 1.32  & 7.19  & 0.58  & 0.06  & 0.06  & 1.40  & 0.06  & 2.80  & 7.19  & 0.73  & 0.58  & 2.05  & 0.30  & 0.56 \\
    \hline
    12    & 0.73  & 0.75  & 1.36  & 7.45  & 0.60  & 0.06  & 0.06  & 1.45  & 0.06  & 2.90  & 7.45  & 0.75  & 0.60  & 1.90  & 0.28  & 0.52 \\
    \hline
    
    13    & 0.71  & 0.74  & 1.34  & 7.35  & 0.59  & 0.06  & 0.06  & 1.43  & 0.06  & 2.86  & 7.35  & 0.74  & 0.59  & 1.96  & 0.33  & 0.43 \\
   \hline
    14    & 0.73  & 0.76  & 1.37  & 7.49  & 0.60  & 0.06  & 0.06  & 1.46  & 0.06  & 2.91  & 7.49  & 0.76  & 0.60  & 1.89  & 0.35  & 0.45 \\
    \hline
    15    & 0.76  & 0.78  & 1.42  & 7.75  & 0.63  & 0.06  & 0.06  & 1.51  & 0.06  & 3.02  & 7.75  & 0.78  & 0.63  & 1.75  & 0.35  & 0.46 \\
    \hline
    16    & 0.73  & 0.76  & 1.37  & 7.52  & 0.61  & 0.06  & 0.06  & 1.46  & 0.06  & 2.93  & 7.52  & 0.76  & 0.61  & 1.87  & 0.35  & 0.46 \\
   \hline
    17    & 0.72  & 0.75  & 1.35  & 7.39  & 0.60  & 0.06  & 0.06  & 1.44  & 0.06  & 2.88  & 7.39  & 0.75  & 0.60  & 1.94  & 0.34  & 0.42 \\
   \hline
    18    & 0.64  & 0.67  & 1.21  & 6.61  & 0.53  & 0.05  & 0.05  & 1.29  & 0.05  & 2.57  & 6.61  & 0.67  & 0.53  & 2.42  & 0.30  & 0.17 \\
   \hline
    19    & 0.72  & 0.74  & 1.35  & 7.36  & 0.59  & 0.06  & 0.06  & 1.43  & 0.06  & 2.86  & 7.36  & 0.74  & 0.59  & 1.95  & 0.33  & 0.33 \\
   \hline
    20    & 0.72  & 0.75  & 1.35  & 7.37  & 0.59  & 0.06  & 0.06  & 1.43  & 0.06  & 2.87  & 7.37  & 0.75  & 0.59  & 1.94  & 0.30  & 0.35 \\
    \hline
    21    & 0.63  & 0.65  & 1.18  & 6.43  & 0.52  & 0.05  & 0.05  & 1.25  & 0.05  & 2.50  & 6.43  & 0.65  & 0.52  & 2.55  & 0.32  & 0.24 \\
    \hline
    22    & 0.67  & 0.70  & 1.26  & 6.91  & 0.56  & 0.05  & 0.05  & 1.34  & 0.05  & 2.69  & 6.91  & 0.70  & 0.56  & 2.20  & 0.32  & 0.24 \\
    \hline
    23    & 0.67  & 0.69  & 1.25  & 6.83  & 0.55  & 0.05  & 0.05  & 1.33  & 0.05  & 2.66  & 6.83  & 0.69  & 0.55  & 2.26  & 0.32  & 0.25 \\
    \hline
    24    & 0.79  & 0.82  & 1.49  & 8.14  & 0.66  & 0.06  & 0.06  & 1.58  & 0.06  & 3.17  & 8.14  & 0.82  & 0.66  & 1.60  & 0.32  & 0.26 \\
   \hline
    25    & 0.65  & 0.67  & 1.21  & 6.63  & 0.54  & 0.05  & 0.05  & 1.29  & 0.05  & 2.58  & 6.63  & 0.67  & 0.54  & 2.39  & 0.32  & 0.27 \\
    \hline
    26    & 0.90  & 0.94  & 1.69  & 9.26  & 0.75  & 0.07  & 0.07  & 1.80  & 0.07  & 3.60  & 9.26  & 0.94  & 0.75  & 1.24  & 0.30  & 0.36 \\
    \hline
    27    & 0.71  & 0.73  & 1.33  & 7.26  & 0.59  & 0.06  & 0.06  & 1.41  & 0.06  & 2.83  & 7.26  & 0.73  & 0.59  & 1.99  & 0.30  & 0.35 \\
    \hline
    28    & 0.67  & 0.70  & 1.26  & 6.88  & 0.56  & 0.05  & 0.05  & 1.34  & 0.05  & 2.68  & 6.88  & 0.70  & 0.56  & 2.23  & 0.32  & 0.27 \\
    \hline
    29    & 0.69  & 0.72  & 1.30  & 7.08  & 0.57  & 0.06  & 0.06  & 1.38  & 0.06  & 2.76  & 7.08  & 0.72  & 0.57  & 2.09  & 0.31  & 0.31 \\
    \hline
    30    & 0.65  & 0.68  & 1.23  & 6.72  & 0.54  & 0.05  & 0.05  & 1.31  & 0.05  & 2.62  & 6.72  & 0.68  & 0.54  & 2.35  & 0.32  & 0.24 \\
    \hline
    31    & 0.60  & 0.62  & 1.12  & 6.14  & 0.50  & 0.05  & 0.05  & 1.19  & 0.05  & 2.39  & 6.14  & 0.62  & 0.50  & 2.81  & 0.33  & 0.21 \\
    \hline
    32    & 0.70  & 0.72  & 1.31  & 7.14  & 0.58  & 0.06  & 0.06  & 1.39  & 0.06  & 2.78  & 7.14  & 0.72  & 0.58  & 2.07  & 0.22  & 0.31 \\
    \hline
    33    & 0.66  & 0.68  & 1.24  & 6.75  & 0.55  & 0.05  & 0.05  & 1.31  & 0.05  & 2.63  & 6.75  & 0.68  & 0.55  & 2.31  & 0.30  & 0.25 \\
    \hline
    34    & 0.71  & 0.74  & 1.34  & 7.32  & 0.59  & 0.06  & 0.06  & 1.42  & 0.06  & 2.85  & 7.32  & 0.74  & 0.59  & 1.99  & 0.30  & 0.26 \\
   \hline
    35    & 0.66  & 0.69  & 1.24  & 6.78  & 0.55  & 0.05  & 0.05  & 1.32  & 0.05  & 2.64  & 6.78  & 0.69  & 0.55  & 2.30  & 0.24  & 0.22 \\
    \bottomrule
    \end{tabular}%
    \captionof{table}{Values of ductility prediction parameters $\Theta$, D parameter for all the 12 crack systems studied and the phenomenological models $\mathrm{D_{surrogate}}$, LLD and Pugh ratio for each HEA.}\label{Phenomenological table}
    \end{table*}
\end{center}%

\subsection{Limitations}

It is worth highlighting that these results are obtained for fully random systems. Experimentally, it is observed that this mixture tends to phase decompose at temperatures lower than the ODTT (order-disorder transition temperature) in WTa rich and CrV rich regions, which might significantly change the ductile/brittle response of the material. In fact, according to our results for solid solutions, the brittleness observed experimentally is likely due to microstructure features induced by second phase precipitation. However, the effect of phase separation is complex and is left for a future study. The bulk structures considered in this work are perfect crystals and do not consider the effects of existing dislocations in the lattice. Note also that manufacturing of these alloys is tremendously challenging and controlling the microstructure far from trivial. In addition, there exist many more crack and slip systems, which have not been studied in this work but can actively contribute to the intrinsic ductility of these alloys and hence would be required for a holistic understanding of the underlying failure mechanisms.

\section{Conclusions} \label{sec:conclusions}

In this study we estimated the intrinsic ductility of WTaCrV HEAs random solutions using DFT calculations relying on multiple ductility prediction models. The validity of the DFT methods, as well as the assumption that the alloys exhibit a BCC structure was tested using experimental data available for $\mathrm{W_{38}Ta_{36}Cr_{15}V_{11}}$. Intrinsic ductility as a function of alloy concentration was studied using the analytical Rice model. In addition, several phenomenological models were analyzed in this work and their potential to improve the efficiency was also discussed. Our key findings are summarized as follows:

\begin{itemize}
  \item The values of the computed elastic constants are compared with the experimental observations for $\mathrm{W_{38}Ta_{36}Cr_{15}V_{11}}$. The small error in our findings as compared to the experimental values validate the robustness of the used DFT framework.

  \item The computed values of the analytical Rice model for a total of 35 WTaCrV HEAs predict intrinsic ductility for 7 out of 12 crack systems studied in this work. These findings hint towards the potential of these alloys to showcase ductile behavior and motivates the experimental validation of these results as future scope of this study.

  \item The isotropy of the HEAs was also studied and the alloys were found to be mainly isotropic in nature using the Zener anisotropy ratio. This is also further validated by computing the anisotropic part of the ductility metric $\chi$, which shows negligible deviation for these alloys for a given crack system.

  \item The results obtained using the analytical model when studied for correlations with the alloying concentrations of the constituent elements show that increasing the concentrations of V and decreasing the concentrations of W can significantly improve the ductility in these HEAs.
  
  \item This work also studies multiple phenomenological ductility models and assesses their potential to improve the computational efficiency of the Rice model by ductility screening. Our results show that the $\mathrm{D_{surrogate}}$ might be a useful parameter due to its strong correlation with the analytical model in isotropic alloys. The Pugh ratio also shows promise when the computation of the elastic constants is easier or desired, although this model is expected to have a lower accuracy. These models are found to be effective ways of screening a large compositional space of HEAs to a more manageable number to be further studied using the more computationally expensive Rice model, thus improving the overall efficiency of the ductility prediction process.

\end{itemize}

\section{Declaration of competing interest}
The authors declare that they have no known competing financial interests or personal relationships that could have appeared to influence the work reported in this paper.

\section{Acknowledgment}
The authors acknowledge the support from the US Department of Energy, Office of Science, Office of Fusion Energy Sciences pilot program under grant number AT2030110.
Clemson University is acknowledged for generous allotment of compute time on the Palmetto cluster. This material is based on work supported by the National Science Foundation under Grant Nos. MRI\# 2024205, MRI\# 1725573, and CRI\# 2010270 \cite{antao2024modernizing}.

\appendix
\label{Appendix A}
\section{Crack propagation in anisotropic media}
This section contains a brief summary of the theoretical derivation for a generalized solution of a crack in a homogeneous linear elastic anisotropic medium. This is based on work by Ting \cite{Ting} and others \cite{Ductile_and_brittle_crack-tip_response,Bower2009-xj,Hwu2010}. The solutions make use of the Eshelby-Reid-Shockley formalism and the Stroh formalism under a 2-dimensional plane strain assumption.

\subsection{Eshelby-Reid-Shockley formalism}\label{appendix A1}
The stress-strain relationship is given by Hooke's law as follows:
\begin{equation}\label{hookes law}
    \mathrm{\sigma_{ij} = C_{ijks}\epsilon_{ks}}
\end{equation}
Where, $\mathrm{\sigma_{ij}}$, $\mathrm{C_{ijks}}$ and $\mathrm{\epsilon_{ks}}$ are the stress matrix, elastic stiffness tensor and strain matrix, respectively. \par
The equilibrium equations are given by:
$\mathrm{\nabla \:.\: \sigma_{ij} + F_{j} = 0}$ , where $\mathrm{\nabla \:.\: \sigma_{ij}}$ is the divergence of the stress matrix and $\mathrm{F_{j}}$ is the resultant force on the material. For equilibrium conditions, $\mathrm{F_{j} = 0}$, we have:
\begin{equation}
\label{equation hookes law differential}
    \mathrm{C_{ijks} u_{k,sj} = 0}
\end{equation}
 \par
We assume that the stiffness matrix has all the elastic symmetries, $\mathrm{C_{ijks} = C_{ksij} = C_{jiks} = C_{jisk} = C_{skji} = C_{ksji} = C_{skij}  }$ \\
For a plane strain assumption, the solution is expected to be a linear equation in $\mathrm{x_{1}}$ and $\mathrm{x_{2}}$. Let z be a linear function of $\mathrm{x_{1}}$ and $\mathrm{x_{2}}$ such that z = $\mathrm{x_{1}}$ + p$\mathrm{x_{2}}$. The displacement field is thus given by $\mathrm{u_{i} = a_{i} f(z)} $ where, $\mathrm{u_{i}}$ is the displacement field, and $\mathrm{a_{i}}$ is an arbitrary vector. Taking the partial derivative of u with respect to s and then with respect to j gives us:
\begin{equation}
    \begin{split}
    \label{equation displacement differential}
    \mathrm{u_{k,s} = (\delta_{s1} + p\delta_{s2})a_{k}f^{'}(z)}\\
    \mathrm{u_{k,sj} = a_{k} (\delta_{s1} + p \delta_{s2})(\delta_{j1} + p \delta_{j2})f^{"}(z)}
    \end{split}
\end{equation}
Using Eqs. \ref{equation displacement differential} and \ref{equation hookes law differential}, we have:
\begin{equation}\label{equation eigen value complicated}
\mathrm{(C_{i1k1} + p(C_{i1k2} + C_{i2k1}) + p^{2}(C_{i2k2}))a_{k} = 0}    
\end{equation} 
From here on, for easy of reading, we will use the following notations:
\begin{equation}\label{Simplified notation}
\begin{split}
\mathrm{Q = C_{i1k1}} \\
\mathrm{R = C_{i1k2}} \\
\mathrm{T = C_{i2k2}} \\
\end{split}
\end{equation}
Hence Eq. \ref{equation eigen value complicated} can be written in simplified notation as $\mathrm{(Q + p(R + R^{T}) + p^{2}T) a_{k} = 0}$. This is an eigenvalue problem. For non-trivial mathematical solutions to this problem, we have the following condition:
\begin{equation}
\label{equation sextic}
    \mathrm{|\: Q + p(R + R^{T}) + p^{2}T \:| = 0}
\end{equation}
Equation \ref{equation sextic} is a sextic equation of a polynomial of degree 3 in $\mathrm{p^{2}}$. This equation will have 3 pairs of complex conjugate roots \cite{Ting}. Let $\mathrm{p_{\alpha}}$ and $\mathrm{a_{\alpha}}$ be the eigenvectors and eigenvalues, respectively, for Eq. \ref{equation sextic}. Without the loss of generality, let Im($\mathrm{p_{\alpha}}$) be positive. This gives the roots a general form as follows:
\begin{equation}
\begin{split}
    \mathrm{p_{\alpha+3} = \bar{p}_{\alpha}}\\
    \mathrm{a_{\alpha+3} = \bar{a}_{\alpha}}\\
\end{split}
\end{equation}
Superimposing all to form a general solution:
\begin{equation}
\mathrm{u = \sum_{\alpha=1}^3 a_{\alpha} f_{\alpha}(z_\alpha) + (\bar{a}_{\alpha}) f_{a+3}(\bar{z}_{\alpha}) }
\end{equation}

\subsection{Stroh formalism}
\label{Stroh formalism}
Equation \ref{equation sextic} can be rearranged and equated to vector b such that:
\begin{equation}
\label{sextic b equation}
    \mathrm{\frac{-(Q + p R) a}{p} = (R^{T} + p T) = b}
\end{equation}
We now define a stress function $\mathrm{\Phi}$ such that:
\begin{equation}
    \mathrm{\Phi = b_i f(z)}
\end{equation}
Also, using Eqs. \ref{hookes law}, \ref{sextic b equation} and \ref{equation displacement differential} we have:
\begin{equation}
    \begin{split}
        \label{stresses}
        \mathrm{\sigma_{i1} = -p b_{i} f^{'}(z)}\\
        \mathrm{\sigma_{i2} = b_{i} f^{'}(z)}
    \end{split}
\end{equation}
and relying on the symmetry of the stress tensor, i.e $\mathrm{\sigma_{12} = \sigma_{21}}$, we have the relation $\mathrm{\Phi_{1,1} + \Phi_{2,2} = 0}$. This gives us the relation:
\begin{equation}
    \mathrm{b_{1} + p b_{2} = 0}
\end{equation}

Writing down the cumulative solutions gives us:
\begin{equation}
\label{displacement and traction}
    \begin{split}
        \mathrm{u = \sum_{\alpha=1}^{3}(a_{\alpha} f_{\alpha}(z_{\alpha})) + (\bar{a}_{\alpha} f_{\alpha+3}(\bar{z}_{\alpha})) }\\
        \mathrm{\Phi = \sum_{\alpha=1}^{3}(b_{\alpha} f_{\alpha}(z_{\alpha})) + (\bar{b}_{\alpha} f_{\alpha+3}(\bar{z}_{\alpha})) }\\
    \end{split}
\end{equation}
For most applications, we will have the same functional form \cite{Ting}. Without loss of generality, we assume the functional form to be as follows:
\begin{equation}
\begin{split}   
    \mathrm{f_{\alpha}(z_{\alpha}) = f_{\alpha}(z_{\alpha}) q_{\alpha}}\\
    \mathrm{f_{\alpha+3}(\bar{z}_{\alpha}) = f(\bar{z}_{\alpha}) \bar{q}_{\alpha}}
    \end{split}
\end{equation}
Hence, the following solutions for u and $\mathrm{\phi}$ can be obtained:
\begin{equation}
\label{displacemet field and stress function}
    \begin{split}
        \mathrm{u = 2 Re (A\langle f(z_{*})\rangle)}\\
        \mathrm{\Phi = 2 Re (B\langle f(z_{*})\rangle)}
    \end{split}
\end{equation}
Where, A and B are stroh matrices A=[$\mathrm{a_{1}, a_{2}, a_{3}}$] and A=[$\mathrm{b_{1}, b_{2}, b_{3}}$], respectively. The function $\mathrm{f(z_{*})}$ = diag[$\mathrm{f(z_{1}), f(z_{2}), f(z_{3})}$] is related to a and b by definition. 

\subsection{A solution for crack tip under uniform loading in a 2-D anisotropic linear elastic medium}
\label{appendix A3}
For plane strain assumptions, we have $\mathrm{\epsilon_{33}}$ and by the defined stress function, we have stresses $\mathrm{\sigma_{i1} = - \phi_{i,2}}$, $\mathrm{\sigma_{i2} =  \phi_{i,1}}$ and $\mathrm{\sigma_{33}}$ from Eq. \ref{hookes law}. For a sharp crack subjected to uniform loading in a homogeneous linear elastic medium, the general solution for a uniform stress $\mathrm{\sigma_{ij}^{\infty}}$ consists of the uniform stress solution for $\mathrm{\sigma_{ij}^{\infty}}$ and a disturbed solution due to the presence of the crack \cite{Ductile_and_brittle_crack-tip_response,Bower2009-xj,Hwu2010}. We look for the singular stress field for the disturbed solution with the crack. \par
Let a crack of length 2a be centrally located at $\mathrm{x_{2} = 0}$ and $\mathrm{|x_{1} < 0|}$. For traction free crack surface, the boundary conditions for the disturbed solution are:
\begin{equation}
    \begin{split}
        \mathrm{\Phi  = 0}, \: for \: \mathrm{|x|} = \infty\\
         \mathrm{\Phi  = -x_{1} t_{\Gamma} }, for \: \mathrm{x_{2} = \pm{0}} \: and \: \mathrm{|x_{1}| < a}
    \end{split}
\end{equation}
The stress vanishes at infinity and a uniform traction $\mathrm{t_{\Gamma} = [\sigma_{21}^{\infty}, \sigma_{22}^{\infty}, \sigma_{23}^{\infty}]^{T}}$ at the upper crack surface and $\mathrm{-t_{\Gamma}}$ to the lower crack surface is applied. Using the equations derived in Eq. \ref{displacemet field and stress function}, we have the stresses:
\begin{equation}
\label{untransformed t}
    \begin{split}
        \mathrm{t_{1} \equiv [\sigma_{11}, \sigma_{12}, \sigma_{13}]^{T} = -Re(B\langle f_{,2}(z_{\alpha})\rangle)B^{-1}}\\
        \mathrm{t_{2} \equiv [\sigma_{21}, \sigma_{22}, \sigma_{23}]^{T} = -Re(B\langle f_{,1}(z_{\alpha})\rangle)B^{-1}}
    \end{split}
\end{equation}
where,\\
$\mathrm{f_{,1}(z_{\alpha}) = \frac{z_{\alpha}}{\sqrt{z_{\alpha}^{2} - a^{2}}} - 1}$\\
$\mathrm{f_{,2}(z_{\alpha}) = \frac{z_{\alpha} p_{\alpha}}{\sqrt{z_{\alpha}^{2} - a^{2}}} - p_{\alpha}}$
\par
The stress intensity factors at crack tips are:\
\begin{equation}
    \mathrm{K = \sqrt{\pi a t_{\Gamma}}}
\end{equation}
Performing a coordinate transformation from the cartesian system $\mathrm{[x_{1}, x_{2}, x_{3}]}$ to the polar cylindrical system [r, $\mathrm{\theta}$, $\mathrm{x_{3}}$] and shifting the origin to the crack tip at (a, 0, 0). This gives us:
\begin{equation}
    \mathrm{z_(\alpha) = r(cos(\theta) + p_{\alpha}sin(\theta))}
\end{equation}

As r $\rightarrow$ 0, we have the asymptotic limits:
\begin{equation}
\label{transformed f}
    \begin{split}
        \mathrm{f_{,1}^{a}(z_{\alpha}) = \frac{\sqrt{a}}{\sqrt{2r(cos(\theta) + p_{\alpha}sin(\theta)}}}\\
        \mathrm{f_{,2}^{a}(z_{\alpha}) = \frac{p_{\alpha}\sqrt{a}}{\sqrt{2r(cos(\theta) + p_{\alpha}sin(\theta)}}}
    \end{split}
\end{equation}
From Eqs. \ref{transformed f} and \ref{untransformed t}, we get the following asymptotic limits:
\begin{equation}
    \label{tractions in polar coordinates}
    \begin{split}
        \mathrm{t_{1}^{a} = -Re(B \langle \frac{p_{\alpha}}{\sqrt{cos(\theta) + p_{\alpha}sin(\theta)}}\rangle B^{-1})K}\\
        \mathrm{t_{2}^{a} = -Re(B \langle \frac{1}{\sqrt{cos(\theta) + p_{\alpha}sin(\theta)}}\rangle B^{-1})K}
    \end{split}
\end{equation}
We seek to determine the angular dependence of the singular stress field in polar coordinates $\mathrm{\sigma_{\theta \alpha}}$ under pure mode $\mathrm{K_{\beta}}$ loading by computing the $\mathrm{F_{\alpha \beta}(\theta)}$ using the Rice model \cite{RICE1992239}. Applying the usual tensor transformation laws of rotation in Eq. \ref{tractions in polar coordinates} about an axis by an angle $\mathrm{\theta}$, one may readily acquire the desired singular stresses in polar coordinates that can then be used to compute $\mathrm{F_{\alpha \beta}(\theta)}$.

\subsection{Computation of $\mathrm{\Lambda}$}
The sextic formalism presented in Eq. \ref{equation sextic} gives out six roots. We assume the material to be degenerate, i.e, six distinct roots and eigenvalues/eigenvectors. Using the notation defined in Eq. \ref{Simplified notation} and the Rice formalism \cite{Ting,Bower2009-xj}, the sextic equation defined previously is rewritten using the fundamental elasticity matrix as an eigen value problem of the form $\textbf{N} \textbf{S} = p \textbf{S}$, where fundamental elasticity matrix $\textbf{N}$ and $\textbf{S}$ are defined as:
\begin{equation}
    \begin{split}
        \mathrm{\textbf{N} = \begin{bmatrix}
N_{1} & N_{2}\\
N_{3} & N_{1}^{T}
\end{bmatrix}}\\
        \mathrm{\textbf{S} = \begin{bmatrix}
a\\
b
\end{bmatrix}}\\
    \end{split}
\end{equation}
Here, $\mathrm{\textbf{N}_{1} = -T^{-1}R^{T}}$, $\mathrm{\textbf{N}_{2} = T^{-1}}$ and $\mathrm{\textbf{N}_{3} = RT^{-1}R^{T} -Q}$. The eigenvalues are the complex conjugate pairs of p and the eigenvectors are the the complex conjugate pairs of $\mathrm{S}$, where $b$ represents the traction part and $a$ represents the displacements from Eq. \ref{displacement and traction}. The Stroh matrices A and B are defined as A = [$\mathrm{a_{1}}$, $\mathrm{a_{2}}$, $\mathrm{a_{2}}$] and B = [$\mathrm{b_{1}}$, $\mathrm{b_{2}}$, $\mathrm{b_{2}}$] respectively. The real value Barnett-Lothe tensor L is defined \cite{Ting,Bower2009-xj,Ductile_and_brittle_crack-tip_response} in the Rice framework as
\begin{equation}
    \begin{split}
        \mathrm{L = -2iBB^{T}} \\
        \mathrm{L =  Re\{-iBA^{-1}}\}
    \end{split}
\end{equation}
The parameter $\mathrm{\Lambda}$ is thus defined as
\begin{equation}
    \mathrm{\Lambda = \frac{L^{-1}}{2}}
\end{equation}

\section{Dependence of simulations on local chemical environment}
\begin{figure}
    \centering
    \includegraphics[width=3.5in]{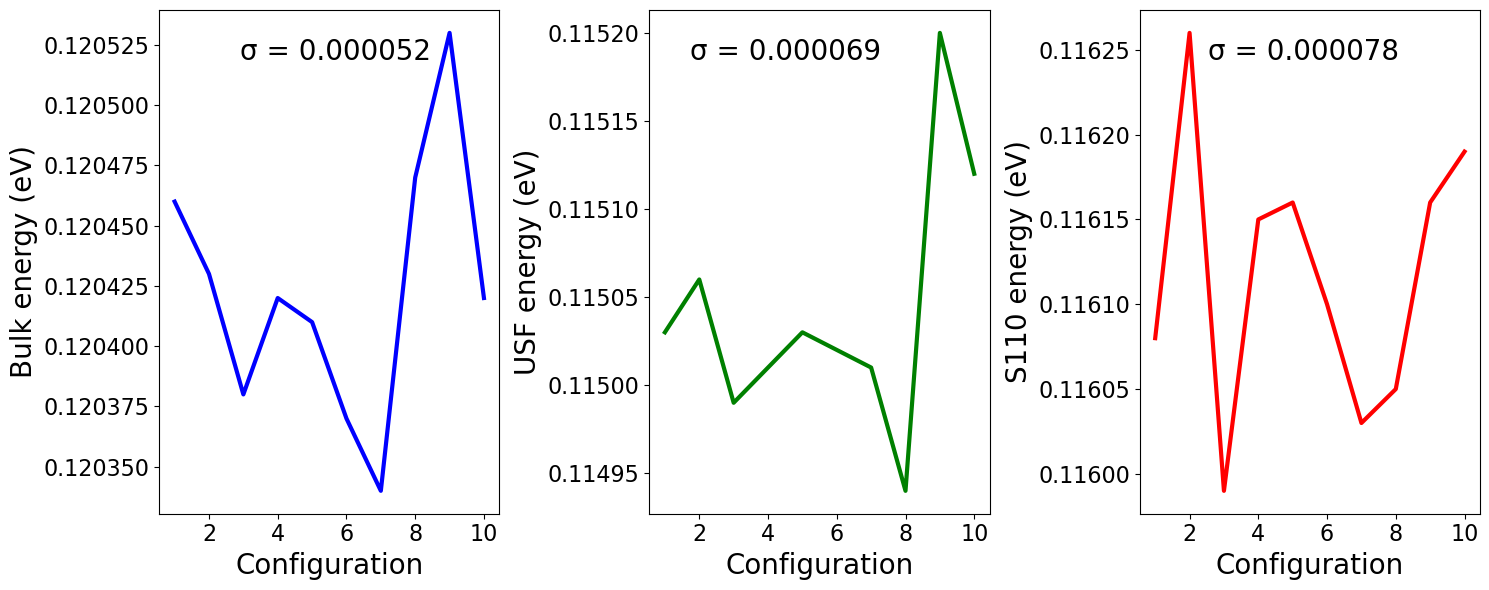}
    \caption{The distribution and standard deviation in the bulk energies, surface energies and stacking fault energies for a crystal of $\mathrm{W_{30}}$$\mathrm{Ta_{30}}$$\mathrm{Cr_{31}}$$\mathrm{V_{9}}$. The Y-axis contains the magnitude of the relaxed energies and the X-axis represents the indices of distinct chemical configurations of the same HEA}
    \label{fig:ds}
\end{figure}

The dependence of our computational simulations on the local chemical environments is discussed. The material properties calculated using DFT are expected to show deviation for each atomic configuration of the same alloy. For the 12 unique crack systems studied in this work, the variation of the bulk energies, $\mathrm{\gamma_{surface}}$ and $\mathrm{\gamma_{usf}}$ with the local chemical environment are studied. The data in Fig. \ref{fig:ds} shows the distribution and standard deviation of the bulk energies, surface energies for the (110) plane and the unstable stacking fault energies in the [111](110) slip system. As discussed in section \ref{Rice theory}, the values for each DFT parameter is taken as the average value of 10 unique random orientations for the same alloy. Our findings show that there is negligible deviation in the values and the standard deviations are small. It has to be noted that for different configurations, only the atomic positions are changed randomly and all the other parameters like cell size, number of atoms and the atomic potentials are kept constant.

\end{document}


\begin{frontmatter}



\title{Supplementary material}


\end{frontmatter}

\section{Phase formation rules for HEAs}
 \label{phase formation in bcc heas}
 During solidification, a typical metallic alloy has three competing states i.e, elemental phase, inter-metallic compounds and solid-solution phase. The Hume-Rothery rules for solid-solution formation are mainly based on alloys with two principal elements. Zhang et. al.\cite{Phase_formation_rules} have presented a HEA specific set of phase formation rules that are used for validating the BCC solid-solution phase assumption in this work. \par
 The comprehensive effect of the atomic size differences is described as follows: 
 \begin{equation}
     \mathrm{\xi = \sum_{i=1} ^{n} \sqrt{\frac{ 1-r_{i}}{\sum_{j=1} ^{n} c_{j}r_{j}}}.}
 \end{equation}
 Here $\mathrm{\delta}$ is the atomic size mismatch, n is the number of elements, $\mathrm{c_{i}}$, $\mathrm{c_{j}}$ are the concentrations and $\mathrm{r_{i}}$, $\mathrm{r_{j}}$ are the atomic radii of the i-th and j-th elements, respectively \cite{Kittel2004}. The rules state that if the atomic size difference between the solute and solvent is more than 14\%, the solubility is likely to be restricted due to the high lattice distortions which will make it less probable to form a solid-solution phase. \par
 The enthalpy of mixing and the entropy of mixing, which are given as:
 \begin{equation}
     \mathrm{\Delta H_{mix} = \sum_{i=1,i\neq j}^{n} \Omega_{ij} c_{i} c_{j}}
 \end{equation}
 \begin{equation}
     \mathrm{\Delta S_{mix} = -R \sum _{i=1}^{n} c_{i} ln c_{i}}
 \end{equation}
 Here, $\Omega_{ij} = 4 \Delta H_{mix}^{binary}$ with $\Delta H_{mix}^{binary}$ the enthalpy of mixing of a binary alloy and R the gas constant. The values of $\Omega_{ij}$ are obtained from the work of Akira et. al.\cite{2005}. As per the phase formation rules, for lower values of $\mathrm{\Delta}$H$\mathrm{_{mix}}$ and higher values of $\mathrm{\Delta}$S$\mathrm{_{mix}}$, the Gibb’s free energy is negative and favours solid-solution formation \cite{Phase_formation_rules}.\par
 The values of VEC for each element is obtained from \cite{Materials_project_10.1063/1.4812323} and the VEC for HEAs is calculated following Vegard's rule of mixtures \cite{vergard}:
 \begin{equation}
     \mathrm{VEC = \sum_{i=1}^{n} c_{i} VEC_{i}}
 \end{equation}
 It has been observed that if the VEC of an alloy is less than 6.87, the solid solution is likely to exhibit a BCC solid solution phase \cite{VEC_10.1063/1.3587228}.

 \subsection{Discussion on phase formation predictions for WTaCrV HEAs}
\begin{figure}
    \centering
    \includegraphics[width=3.4in]{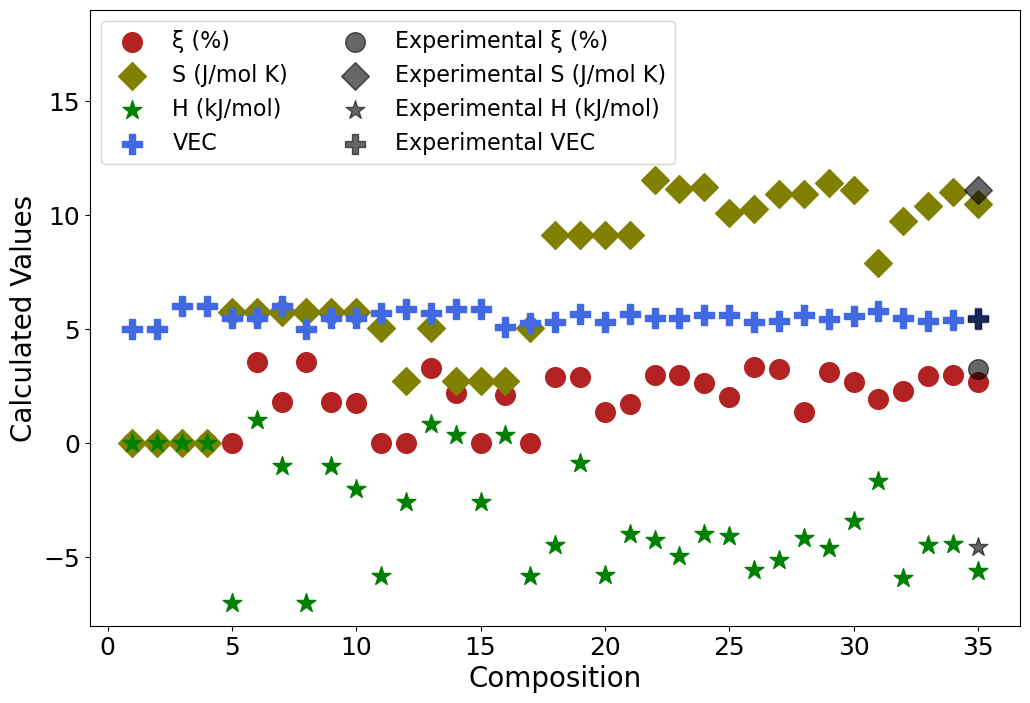}
    \caption{The theoretical phase formation parameters for HEAs along with the experimental values of these parameters for HEA $\mathrm{W_{37}Ta_{37}Cr_{15}V_{11}}$ (index 35). }
    \label{ecexp}
\end{figure}

A comprehensive evaluation of the phase formation potential for these WTaCrV HEAs to form BCC solid solution is discussed using parameters such as the atomic radii mismatch ($\mathrm{\xi}$), mixing enthalpies ($\mathrm{\Delta H_{mix}}$), mixing entropies  ($\mathrm{\Delta S_{mix}}$) and the VEC. Table \ref{Table ec} is a tabulated representation of the parameters used to study the phase formation for the various HEA compositions. This data is cross-validated using experimental data for $\mathrm{W_{38}Ta_{36}Cr_{15}V_{11}}$ \cite{doi:10.1126/sciadv.aav2002,TUNES2023101796}. Figure \ref{ecexp} shows the theoretical and experimental values for the phase formation parameters for WTaCrV HEAs. Following Ref. \cite{Phase_formation_rules}, the enthalpies of mixing should lie within the range of -15 kJmol$^{-1}$ to 5 kJmol$^{-1}$ and the entropies of mixing should be in the range of 12 kJK$^{-1}$mol$^{-1}$ to 17.5 kJK$^{-1}$mol$^{-1}$ in order to form solid solutions for quarternaty alloys. A similar reference for $\mathrm{\xi}$ is the condition that $\mathrm{\xi} < 6.6\%$ for the formation of solid solutions. Also, the VEC values for each HEA are used as predictors for phase formation, where BCC phases are favored for VEC$<6.87$ \cite{TUNES2023101796}. This is observed in our findings as well, indicating the potential to form BCC alloys over face-centered cubic (FCC). Our findings show that the $\mathrm{\Delta H_{mix}}$ and $\mathrm{\xi}$ values are also well within the expected range to favor the formation of stable solid solutions. However, $\mathrm{\Delta S_{mix}}$ is lower than expected to form solid solutions. Also, it is to be noted that the results for the phase formation parameters $\mathrm{\xi}$, $\mathrm{\Delta S_{mix}}$, $\mathrm{\Delta H_{mix}}$ and VEC hold well against experimental data, hence validating the robustness of the framework.

\begin{center}
\begin{table}[htbp]    
    \begin{tabular}{|c|c|c|c|c|c|}
    \toprule
    \thead{Index \\ number} &    
    \thead{HEA} & 
    \thead{$\mathrm{\xi}$ \\ (\%)} & 
    \thead{$\mathrm{\Delta S_{mix}}$ \\  (kJ/mol)} & 
    \thead{$\mathrm{\Delta H_{mix}}$ \\  (kJ/mol)} & 
    \thead{VEC}  \\
    \midrule
     1 & $\mathrm{Ta}$ & 0.00&0.00&0.00&5.00  \\
     2 & $\mathrm{Cr}$ & 0.00&0.00&0.00&5.00 \\
     3 & $\mathrm{V}$ & 0.00&0.00&0.00&6.00  \\
     4 & $\mathrm{W}$ & 0.00&0.00&0.00&6.00  \\
     5 & $\mathrm{W_{50}}$$\mathrm{Ta_{50}}$ &0.00&5.76&-7.00&5.50  \\
     6 & $\mathrm{W_{50}}$$\mathrm{Cr_{50}}$ & 3.57& 5.76& 1.00 & 5.50 \\
     7 & $\mathrm{W_{50}}$$\mathrm{V_{50}}$ & 1.81& 5.76& -1.00&6.00 \\
     8 & $\mathrm{Ta_{50}}$$\mathrm{Cr_{50}}$ & 3.57& 5.76& -7.00& 5.00 \\
     9 & $\mathrm{Ta_{50}}$$\mathrm{V_{50}}$ & 1.81& 5.76& -1.00& 5.50 \\
     10 & $\mathrm{Cr_{50}}$$\mathrm{V_{50}}$ & 1.75& 5.76& -2.00& 5.50 \\
     11 & $\mathrm{W_{70}}$$\mathrm{Ta_{30}}$ & 0.00& 5.05& -5.84& 5.70\\
     12 & $\mathrm{W_{90}}$$\mathrm{Ta_{10}}$ & 0.00& 2.73& -2.56& 5.90 \\
     13 & $\mathrm{W_{70}}$$\mathrm{Cr_{30}}$ & 3.31& 5.05& 0.83& 5.70\\
     14 & $\mathrm{W_{90}}$$\mathrm{Cr_{10}}$ & 2.22& 2.73& 0.36& 5.90 \\
     15 & $\mathrm{W_{90}}$$\mathrm{Ta_{10}}$ &0.00& 2.73& -2.56& 5.90 \\
     16 & $\mathrm{W_{10}}$$\mathrm{Cr_{90}}$ & 2.10&2.73&0.36&5.10 \\
     17 & $\mathrm{W_{30}}$$\mathrm{Ta_{70}}$ & 0.00& 5.05& -5.83& 5.29\\
     18 & $\mathrm{Ta_{33}}$$\mathrm{Cr_{33}}$$\mathrm{V_{33}}$ & 2.92& 9.13&-4.44&5.33 \\
     19 & $\mathrm{W_{33}}$$\mathrm{Cr_{33}}$$\mathrm{V_{33}}$ &2.91&9.13&-0.88&5.67 \\
     20 & $\mathrm{W_{33}}$$\mathrm{Ta_{33}}$$\mathrm{Cr_{33}}$&1.35&9.13&-5.77&5.33 \\
     21 & $\mathrm{W_{33}}$$\mathrm{Ta_{33}}$$\mathrm{V_{33}}$& 1.72&9.13&-4.00&5.66 \\
     22 & $\mathrm{W_{25}}$$\mathrm{Ta_{25}}$$\mathrm{Cr_{25}}$$\mathrm{V_{25}}$ &2.98& 11.52& -4.25& 5.50 \\
     23 & $\mathrm{W_{34}}$$\mathrm{Ta_{30}}$$\mathrm{Cr_{21}}$$\mathrm{V_{15}}$ & 2.97& 11.13& -4.95& 5.49 \\
     24 & $\mathrm{W_{30}}$$\mathrm{Ta_{25}}$$\mathrm{Cr_{15}}$$\mathrm{V_{31}}$ &2.65& 11.24& -3.96& 5.60 \\
     25 & $\mathrm{W_{31}}$$\mathrm{Ta_{32}}$$\mathrm{Cr_{5}}$$\mathrm{V_{32}}$ &2.04& 10.11& -4.07& 5.64 \\
     26 & $\mathrm{W_{31}}$$\mathrm{Ta_{32}}$$\mathrm{Cr_{31}}$$\mathrm{V_{6}}$ &3.33& 10.27& -5.55& 5.31 \\
     27 & $\mathrm{W_{30}}$$\mathrm{Ta_{30}}$$\mathrm{Cr_{31}}$$\mathrm{V_{9}}$ & 3.26& 10.94& -5.12& 5.36\\
     28 & $\mathrm{W_{30}}$$\mathrm{Ta_{30}}$$\mathrm{Cr_{9}}$$\mathrm{V_{31}}$& 1.35& 10.94&-4.15&5.60\\
     29 & $\mathrm{W_{28}}$$\mathrm{Ta_{27}}$$\mathrm{Cr_{27}}$$\mathrm{V_{18}}$ & 3.10& 11.42& -4.61& 5.46 \\
     30 & $\mathrm{W_{19}}$$\mathrm{Ta_{20}}$$\mathrm{Cr_{20}}$$\mathrm{V_{41}}$ & 2.70& 11.08& -3.39& 5.59  \\
     31 & $\mathrm{W_{10}}$$\mathrm{Ta_{10}}$$\mathrm{Cr_{10}}$$\mathrm{V_{70}}$ & 1.95& 7.90& -1.67& 5.79 \\
     32 & $\mathrm{W_{41}}$$\mathrm{Ta_{41}}$$\mathrm{Cr_{9}}$$\mathrm{V_{9}}$ & 2.27& 9.74& -5.92& 5.50 \\
     33 & $\mathrm{W_{7}}$$\mathrm{Ta_{31}}$$\mathrm{Cr_{31}}$$\mathrm{V_{31}}$ & 2.95& 10.41& -4.45& 5.37\\
     34 & $\mathrm{W_{10}}$$\mathrm{Ta_{30}}$$\mathrm{Cr_{30}}$$\mathrm{V_{30}}$ &2.98& 11.02& -4.43& 5.40\\
     35 & $\mathrm{W_{38}}$$\mathrm{Ta_{36}}$$\mathrm{Cr_{15}}$$\mathrm{V_{11}}$ & 2.67& 10.50& -5.61& 5.48 \\
    \bottomrule
    \end{tabular}
    \label{Table 1}
    \captionof{table}{Difference in atomic radii mismatch ($\mathrm{\Delta}$), mixing enthalpies ($\mathrm{\Delta H_{mix}}$), mixing entropies  ($\mathrm{\Delta S_{mix}}$) and the valence electron concentrations (VEC) for WTaCrV HEAs. The indexes in column 1 refer to the unique HEAs in column throughout this work. }%
\end{table}
\end{center}
\label{Table ec}

\label{phase formation label}

\section{Ductility as a function of alloying concentration}

\begin{figure*}
  \includegraphics[width=19cm]{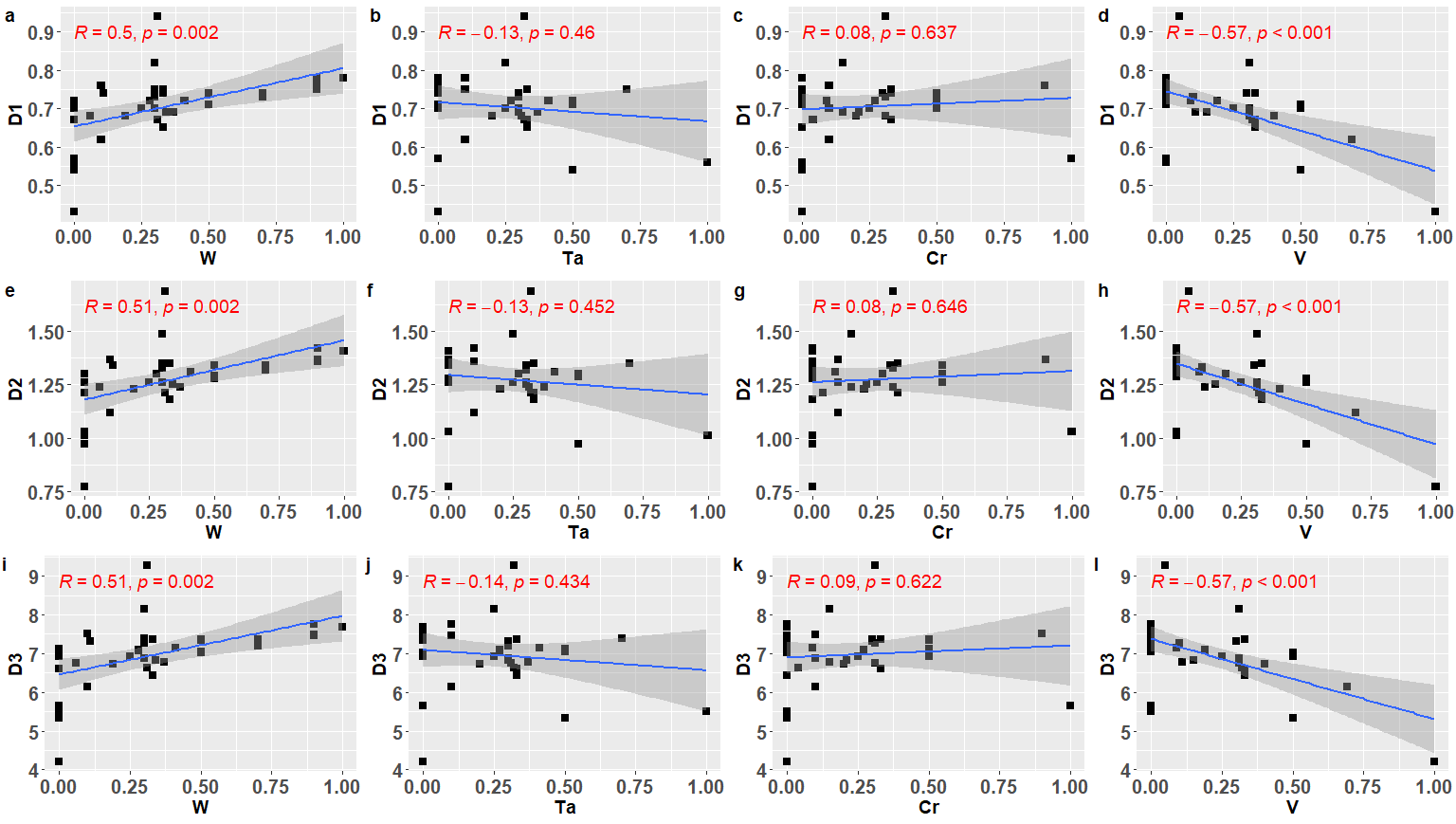}
  \includegraphics[width=19cm]{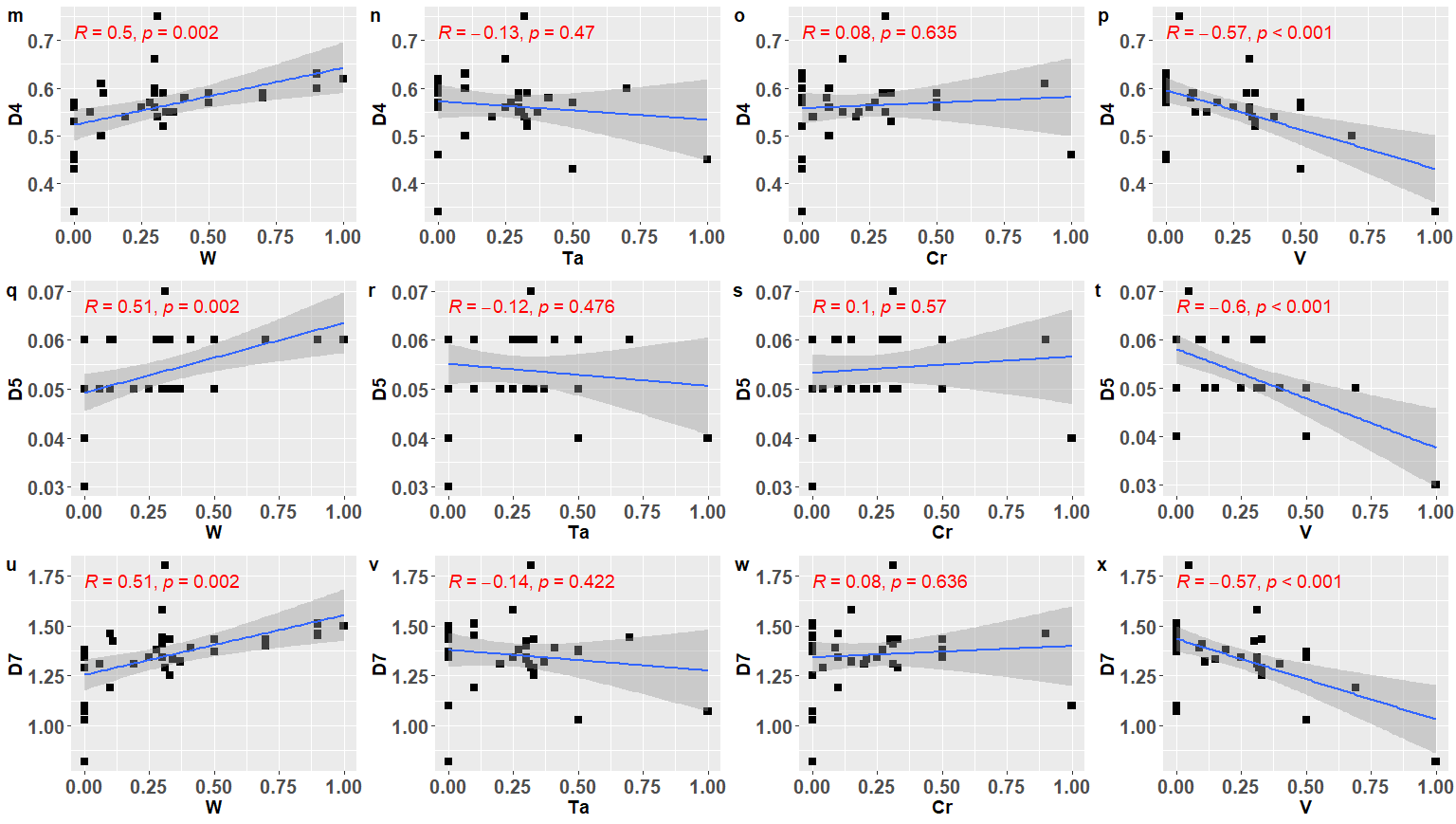}
  \caption{$\mathrm{\textbf{Correlations of the D parameter for the crack systems 1-12 vs the concentrations of alloying elements.}}$ Since some of the systems are identical numerically (see table \ref{table chi}), only crack systems 1, 2, 3, 4, 5 and 7 are shown for avoiding redundancies.}
  \label{D element png}
\end{figure*}

For the sake of succinctness and to eliminate redundancies, the ductility predictions using the Rice model as a function of the alloying elements is not shown for all the 12 geometries in the main text. Hence the results for the remaining geometries are shown here in figure \ref{D element png}. The trends in correlation are similar to the trends we see in the main text which further bolsters the reliability of the correlation.

\section{Correlation between the Rice model and phenomenological}

\begin{figure*}
  \includegraphics[width=19cm]{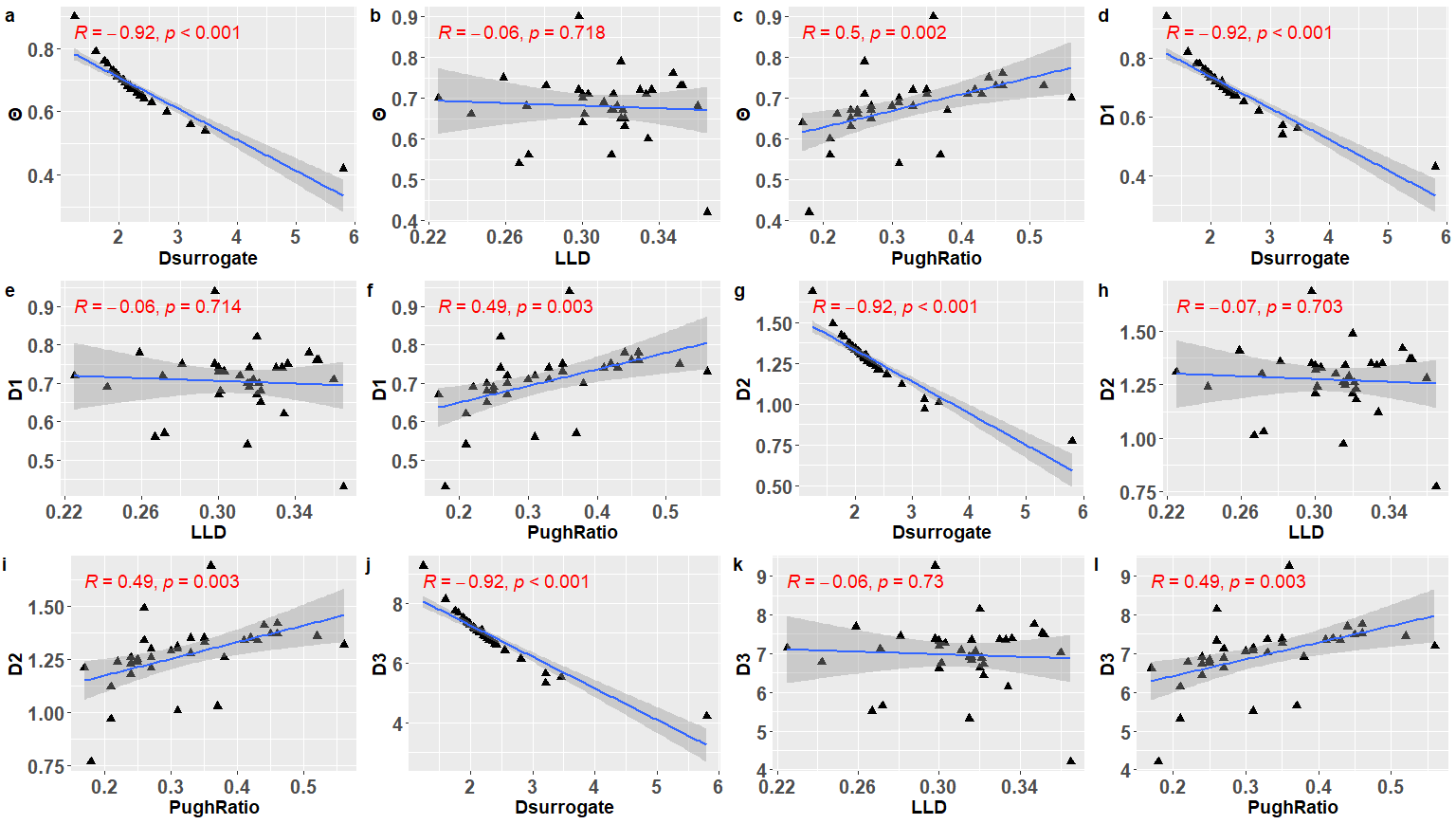}
  \includegraphics[width=19cm]{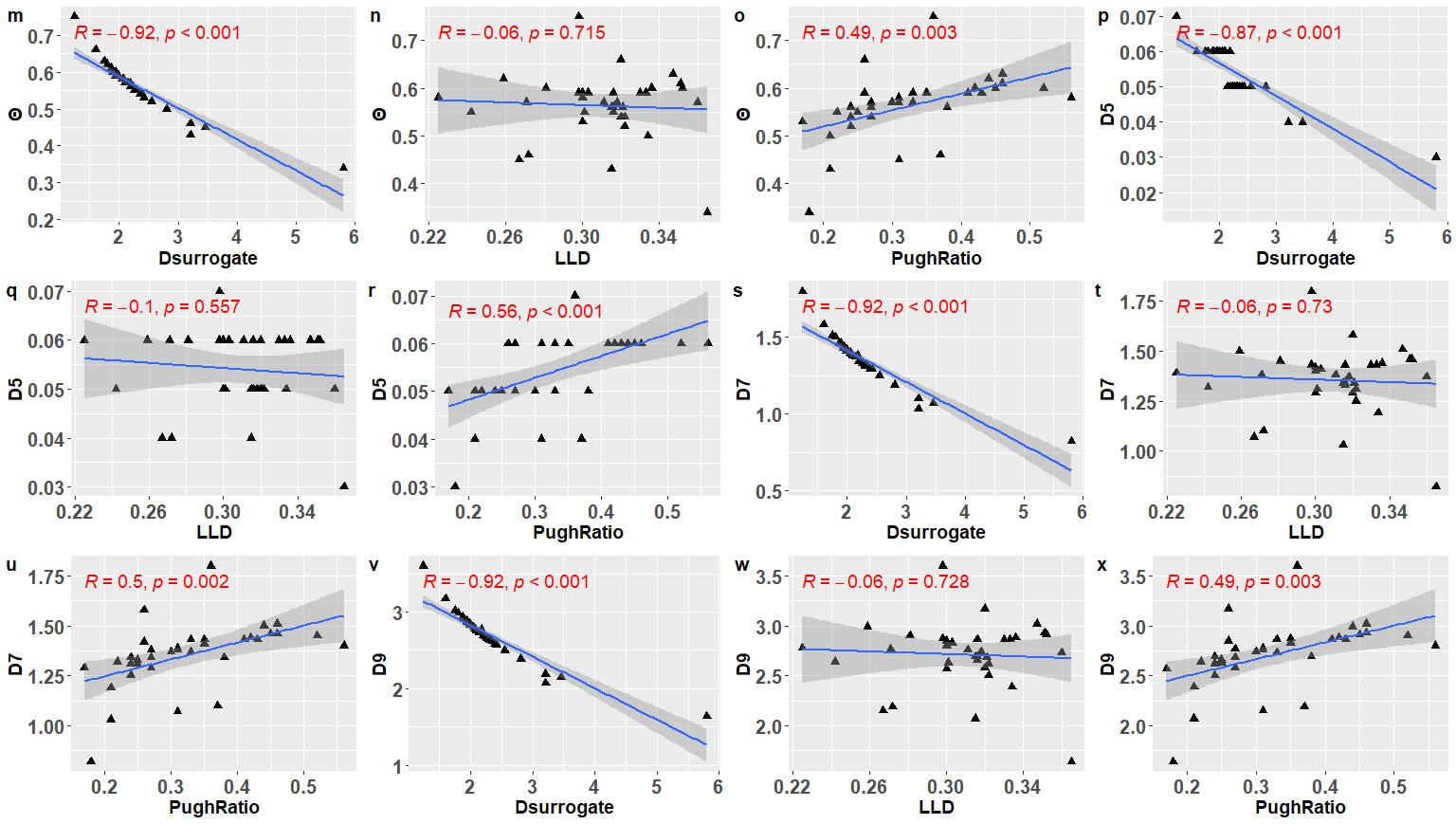}
  \caption{$\mathrm{\textbf{Correlations of the Phenomenological models with the analytical D parameter.}}$ \textbf{(a)} $\mathrm{\Theta}$ vs $\mathrm{D_{surrogate}}$. \textbf{(b)} $\mathrm{\Theta}$ vs LLD. \textbf{(c)}  $\mathrm{\Theta}$ vs LLD. \textbf{(d)}-(\textbf{x}) The correlation of the phenomenological models with the Rice model for crack systems 1, 2, 3, 5, 7 and 9. Since some of the systems are identical numerically (see table \ref{table chi}), only select crack systems are shown for avoiding redundancies.}\label{D geometry png}
\end{figure*}

Similar to the previous section, the main paper leaves out some geometries which can be found in figure \ref{D geometry png}. Once again the trends indicate the exact conclusion that is drawn in the main text.

